\documentclass{pos1}

\usepackage{amssymb}
\usepackage{bm}
\usepackage[numbers,square,sort&compress]{natbib}

\newcommand{\beq}{\begin{equation}}
\newcommand{\eeq}{\end{equation}}
\newcommand{\req}[1]{Eq.~(\ref{#1})}
\newcommand{\aB}{a_\mathrm{B}}
\newcommand{\aion}{a_\mathrm{i}}
\newcommand{\am}{a_\mathrm{m}}
\newcommand{\ChiH}{\bm{\chi}^\mathrm{H}}
\newcommand{\ChiA}{\bm{\chi}^\mathrm{A}}

\newcommand{\dd}{\mathrm{d}}
\newcommand{\EF}{\epsilon_\mathrm{F}}
\newcommand{\gcc}{\mbox{g~cm$^{-3}$}}
\newcommand{\kB}{k_\mathrm{B}}
\newcommand{\khat}{\hat{\mathbf{k}}}
\newcommand{\Kp}{K_\perp}
\newcommand{\mel}{m_\mathrm{e}}
\newcommand{\mion}{m_\mathrm{i}}
\newcommand{\nel}{n_\mathrm{e}}
\newcommand{\nion}{n_\mathrm{i}}
\newcommand{\omc}{\omega_\mathrm{c}}
\newcommand{\omci}{\omega_\mathrm{ci}}
\newcommand{\Omk}[1]{\Omega_{#1}}
\newcommand{\ompe}{\omega_{\mathrm{pe}}}
\newcommand{\opac}{\varkappa}
\newcommand{\rhos}{\rho_\mathrm{s}}
\newcommand{\sSB}{\sigma_\mathrm{SB}}
\newcommand{\Tc}{T_\mathrm{crit}}
\newcommand{\Teff}{T_\mathrm{eff}}
\newcommand{\Ts}{T_\mathrm{s}}
\newcommand{\ycol}{y_\mathrm{col}}
\newcommand{\zete}{\zeta_\mathrm{e}}
\newcommand{\zeti}{\zeta_\mathrm{i}}

\title{Atmospheres and radiating surfaces of neutron stars 
with strong magnetic fields}

\ShortTitle{Magnetized atmospheres of neutron stars}

\author{\speaker{Alexander Y. Potekhin},$^{abcd}$
   Wynn C. G. Ho,$^{e}$
   Gilles Chabrier$^{\,df}$
\\
\llap{$^a$}Ioffe Institute,
      Politekhnicheskaya 26, Saint Petersburg, 194021,
       Russia\\
\llap{$^b$}St Petersburg Polytechnic University,
      Politeknicheskaya 29, Saint Petersburg, 195251,
       Russia\\
\llap{$^c$}Central Astronomical Observatory at Pulkovo,
      Pulkovskoe Shosse 65, Saint Petersburg, 196140,
       Russia\\
\llap{$^d$}Centre de Recherche Astrophysique de Lyon, Universit\'e de
Lyon, Universit\'e Lyon 1, Observatoire de Lyon, Ecole Normale
Sup\'erieure de Lyon, CNRS, UMR 5574, 46 all\'ee d'Italie,
69364, Lyon Cedex 07, France\\
\llap{$^e$}Mathematical Sciences, Physics \& Astronomy, and STAG Research
Centre, University of Southampton, Southampton,
SO17 1BJ, UK\\
\llap{$^f$}School of Physics, University of Exeter, Exeter, EX4 4QL, UK
\\
E-mail: \email{palex@astro.ioffe.ru},
 \email{wynnho@slac.stanford.edu},
 \email{gilles.chabrier@ens-lyon.fr}
}

\abstract{We review the current status of the theory of thermal
emission from the surface layers of neutron stars with strong magnetic
fields $B\sim 10^{10}-10^{15}$~G, including formation of the spectrum
in a partially ionized atmosphere and at a condensed surface.
In particular, we describe recent
progress in modeling partially ionized atmospheres of central compact
objects in supernova remnants, which may have moderately strong fields
$B\sim 10^{10}-10^{11}$~G.  Special attention is given to
polarization of thermal radiation emitted by a neutron star
surface. Finally, we briefly describe applications of the theory to
observations of thermally emitting isolated neutron stars.
}

\FullConference{The Modern Physics of Compact Stars 2015\\
		30 September 2015 - 3 October 2015\\
		Yerevan, Armenia}

\begin{document}

\section{Introduction}

One of the first expectations of neutron-star astrophysics is the
possibility to detect thermal radiation from their hot surfaces, whose
temperature $T \sim10^6$~K implies the maximum of the thermal spectrum
in the range of hundreds eV \cite{Tsuruta64,Chiu64}. However, the
first discovered X-ray sources beyond the Solar system
\cite{Giacconi_ea62,Bowyer_ea64a,Bowyer_ea64b} appeared to be
unrelated with the neutron star radiation \cite{Bowyer_ea64b}. After
Sco X-1, the first discovered cosmic X-ray source
\cite{Giacconi_ea62}, had also been detected in the optical band
\cite{Sandage_ea66}, Shklovsky \cite{Shklovsky67} argued that its
X-ray radiation originated from the accretion of matter onto a neutron
star from its companion. This conjecture was initially refuted
\cite{CameronScoX1}, but later proved to be true \cite{deFreitas77}.
However, the first detection of thermal radiation from the
\emph{surface} of isolated neutron stars (INSs)
\cite{ChengHelfand83,BrinkmannOgelman87} had to wait the launch of
focusing X-ray telescopes \textit{Einstein} (1978--1981) and
\textit{EXOSAT} (1983--1986), which allowed for a dramatic increase in
sensitivity to faint point X-ray sources. The next significant advance
was due to the \textit{ROSAT} X-ray observatory (1990--1998),
which was the first to reliably identify spectra of the thermal
radiation from several pulsars
\cite{Ogelman95,Becker99}. In the 21st century, the data collected by
the new generation X-ray observatories \textit{Chandra} and
\textit{XMM-Newton} give a further impetus to the studies of thermal
radiation of neutron stars.

One of the main challenges in the analysis of the neutron star spectra
is to disentangle different emission components, overlapping in the
X-ray energy range. After discriminating thermal emission, a detailed
study of the  thermal spectra can yield precious information about the
neutron star surface composition and temperature and magnetic field
distributions, the properties of dense
magnetized plasmas in their envelopes and atmospheres, and eventually
constraint the equation of state (EOS) of the ultradense matter in the
neutron star cores. 

The number of known neutron stars with an unambiguously identified
thermal component in the spectrum is not large, but it steadily
increases. Some of them can be understood with models of non-magnetic
atmospheres, whereas others are believed to be endowed with strong
magnetic fields. The non-magnetic neutron-star
atmospheres have been studied in many works
(see Sect.~\ref{sec:models} for magnetic model spectra).
For the atmospheres composed of hydrogen, databases of model spectra 
have been published (models \textsc{nsagrav} \cite{ZavlinPS96},
\textsc{nsspec} \cite{GaensickeBR02}, and \textsc{nsatmos}
\cite{Heinke_ea06} in the database \textit{XSPEC} \cite{XSPEC}), and a
computer code for their calculation has been released
\cite{Haakonsen_ea12}. Model spectra were also calculated for
different chemical compositions
\cite{GaensickeBR02,RajagopalRomani96,Pons_ea02,Heinke_ea06,HoHeinke09,Suleimanov_ea14}
and mixtures \cite{GaensickeBR02,Pons_ea02}. Databases of model
spectra have been published for the atmospheres
composed of iron and of the solar mixture of elements (\textsc{nsspec}
\cite{GaensickeBR02}), of helium (\textsc{nsx} \cite{HoHeinke09}), and
of carbon (\textsc{nsx} \cite{HoHeinke09} and \textsc{carbatm}
\cite{Suleimanov_ea14}). Examples of thermal spectra 
successfully interpreted with the non-magnetic atmosphere models
include bursters, soft X-ray
transients in quiescence, some central compact objects (CCOs) in
supernova remnants, and some millisecond pulsars (see the table in
the review \cite{P14}). Measuring the basic parameters of neutron
stars using model non-magnetic atmospheres has been also recently
reviewed in \cite{Suleimanov_ea16}.

Here, however, we will focus exclusively on models of
those thermal spectra of the INSs that are significantly affected by
strong magnetic fields
(see \cite{P14} for more detailed discussion).
The paper is organized as follows. In
Sect.~\ref{sec:theory} we overview the theory of partially ionized
neutron star atmospheres with strong magnetic fields. In
Sect.~\ref{sec:cond} we describe the model of a condensed
radiating surface and hybrid models of a condensed surface covered by a thin
atmosphere. Section~\ref{sec:pol} is devoted to the polarization of
the neutron star thermal radiation that may be measured in future.
Available examples of successful applications of the theory to
observations are overviewed in Sect.~\ref{sec:obs}, and conclusions
are given in Sect.~\ref{sec:concl}.

\section{Theory of strongly magnetized neutron-star atmospheres}
\label{sec:theory}

We call an atmosphere strongly magnetized, if the magnetic field $B$
strongly (non-perturbatively) affects its opacities and radiative
transfer for thermal photons. This is the case if the electron
cyclotron energy $ \hbar\omc \approx 11.577\,B_{12}$ keV is
greater than either the photon energies $\hbar\omega$ or the atomic
binding energies, or both. Here,
$\omega$ is the photon angular frequency, $\omc=eB/\mel c$ is the
electron cyclotron frequency, $\mel$ and $-e$ are the electron mass
and charge, $c$ is the speed of light, and $B_{12}\equiv
B/(10^{12}\mbox{~G})$.
These two conditions imply
\beq
 B\gtrsim \frac{\mel c }{ \hbar e}\,\kB T \sim 10^{10}\,
    T_6\mbox{~G},
\quad
  B\gtrsim
   B_0 = \frac{\mel^2\,c\,e^3}{\hbar^3} =
   2.3505\times10^9\mbox{~G},
\label{B0}
\eeq
where $T_6 = T/10^6$~K.
The quantity $B_0$ in \req{B0} determines the atomic magnetic-field
parameter $\gamma\equiv B/B_0$. It is also convenient to define the
relativistic  magnetic-field parameter $b\equiv\hbar\omc/\mel c^2 =
B_{12}/44.14$. We call magnetic field superstrong if $b
\gtrsim 1$. In this case the specific effects of Quantum
Electrodynamics (QED) become quite important. The superstrong fields
are believed to exist at the surfaces of magnetars and high-$B$
pulsars (see the
reviews \cite{OlausenKaspi,Mereghetti_ea15,Turolla_ea15}).

\subsection{Radiative transfer in normal modes}
\label{sec:RT}

As has been demonstrated in \cite{GP73}, at typical conditions in
neutron star photospheres one can describe radiative transfer in
terms of specific intensities of two normal polarization modes
\cite{Ginzburg}. This is true if the magnetic field is strong and
$\omega$ lies outside narrow
frequency ranges near resonances and above the electron plasma
frequency $ \ompe=\left({4\pi e^2 \nel / \mel } \right)^{1/2}$, where
$\nel$ is the electron number density. The normal modes, called
extraordinary (X-mode, denoted by subscript or superscript $j=1$ or
X) and ordinary (O-mode, $j=2$ or O), have different polarization vectors
$\bm{e}_j$ and different absorption and scattering coefficients, which
depend on the angle $\theta_{kB}$ between the wave vector $\bm{k}$ and
the magnetic field $\bm{B}$. The radiative transfer equations for the
two normal modes read \cite{KaminkerPS82}
\beq
 \cos\theta_k \frac{\dd I_{\omega,j}(\khat)}{\dd \ycol} =
    \opac_{\omega,j}(\khat) I_{\omega,j}(\khat) 
-
 \frac12\,\opac_{\omega,j}^\mathrm{a}(\khat)
     \mathcal{B}_{\omega,T}
-
      \sum_{j'=1}^2 \int_{(4\pi)} 
        \opac_{\omega,j'j}^\mathrm{s}(\khat',\khat)
         I_{\omega,j'}(\khat') \,\dd\Omk{\bm{k}'},
\label{RTEmag}
\eeq
where $I_{\omega,j}$ denotes the specific intensity  of the
polarization mode $j$ per unit circular frequency (if $I_\nu$ is the
specific intensity per unit frequency, then $I_\omega=I_\nu/(2\pi)$
\cite{Zheleznyakov}), $\khat$ is the unit vector along the wave vector
$\bm{k}$,
\beq
\mathcal{B}_{\omega,T} = 
\frac{\hbar\omega^3}{4\pi^3c^2}
\left(\mathrm{e}^{\hbar\omega/ \kB T}-1\right)^{-1}
\label{Planck}
\eeq
is the specific intensity of non-polarized blackbody
radiation,
\beq
\opac_{\omega,j}(\khat) \equiv
\opac_{\omega,j}^\mathrm{a}(\khat) + \sum_{j'=1}^2
\int_{(4\pi)}
\opac_{\omega,j'j}^\mathrm{s}(\khat',\khat)\,\dd\Omk{\bm{k}'}
\eeq
is the total opacity, $\opac_{\omega,j}^\mathrm{a}$ and
$\opac_{\omega,j'j}^\mathrm{s}(\khat',\khat)$ are its components due
to, respectively, the true absorption and the scattering that changes
the ray direction from $\khat'$ to $\khat$ and may also change the
polarization from $j'$ to $j$, $\dd\Omk{\bm{k}} \equiv
\sin\theta_{k}\dd\theta_{k}\dd\varphi_{k}$ is a solid angle element,
$\theta_k$ is the angle between $\bm{k}$ and the normal to the
surface $\mathbf{n}$, and $\varphi_k$ is the angle in the surface
plane. Finally,
$\ycol = \int_r^\infty (1+z_g)\,\rho(r)\dd r$ is the column density,
where the factor $1+z_g= (1-x_g)^{-1/2}$ is the scale change
due to the gravity according to the General Relativity (GR), $z_g$ is
the gravitational redshift, 
$x_g=2{GM}/{c^2 R}\approx(M/M_\odot)\,(\mbox{2.95 km}/R)$
is the compactness parameter
of the star with mass $M$ and radius $R$, and $G$ is the gravitational
constant (for a typical neutron star $x_g$ lies between 1/5 and 1/2).
The radiative transfer equations should be supplemented with the
hydrostatic and energy balance equations (see
\cite{SuleimanovPouW12}).

The major axis of the polarization ellipse is transverse to
$\bm{B}$ for the X-mode and coplanar with $\bm{B}$ for the
O-mode. Since the magnetic field hampers electron motion in the
transverse direction, it also suppresses 
the damping of the X-wave by the electrons, making
the opacities of the X-mode strongly reduced, if $\omc\gg\omega$.
The dependence of the opacities on $\khat$ and $\khat'$ is also
affected by $\bm{B}$. Therefore, the emission of a magnetized
atmosphere depends on the angles $\theta_k$, $\varphi_k$, and the
angle $\theta_{nB}$ between $\bm{B}$ and 
$\mathbf{n}$, unlike in the non-magnetic case where it depends
only on $\theta_k$. In practice, one usually neglects the dependence
of scattering on the angle between $\khat$ and $\khat'$, in which case
(called the approximation of isotropic scattering) 
$\opac_{\omega,j'j}^\mathrm{s}(\khat',\khat)$ depends only
on the angles $\theta_{kB}$ and $\theta_{kB}'$ that the wave vectors
$\bm{k}$ and $\bm{k}'$ make with $\bm{B}$.

The polarization vectors of normal modes $\bm{e}_{\omega,j}$ are
determined by the complex dielectric tensor $\bm{\varepsilon}(\omega)$
and magnetic permeability tensor. The latter can be set
equal to the unit tensor $\mathbf{I}$ for a gaseous plasma (see
\cite{Ginzburg}). In the Cartesian system with unit vectors
$\hat{\mathbf{e}}_{x,y,z}$ such that $\hat{\mathbf{e}}_{z}$ is
directed along $\bm{B}$,
\begin{equation}
  \bm{\varepsilon} = \mathbf{I} + 4\pi\bm{\chi}
          = 
 \left( \begin{array}{ccc}
 \varepsilon_\perp & \mathrm{i} \varepsilon_\wedge & 0 \\
 -\mathrm{i}\varepsilon_\wedge & \varepsilon_\perp & 0 \\
 0 & 0 & \varepsilon_\| 
 \end{array} \right),
\label{eps-p}
\end{equation}
where
$\bm{\chi}=\ChiH+\mathrm{i}\ChiA$ is the complex polarizability tensor
of plasma, $\ChiH$ and $\ChiA$ are its Hermitian and anti-Hermitian
parts, respectively. The anti-Hermitian part $\ChiA(\omega)$ is
determined by the absorption opacities, and the Hermitian part
$\ChiH(\omega)$ can be obtained from $\ChiA(\omega)$ using the
Kramers-Kronig relation \cite{BulikPavlov,KK}.

In strong magnetic fields, specific QED effects called polarization
and magnetization of vacuum can be important (see, e.g.,
\cite{PavlovGnedin}). At $B\lesssim 10^{16}$, the vacuum polarization
is weak. Then
the total dielectric tensor is $ \bm{\varepsilon} = \mathbf{I} +
4\pi\bm{\chi}(\omega) + 4\pi\bm{\chi}^\mathrm{vac},$ where the vacuum
polarizability tensor $\bm{\chi}^\mathrm{vac}$ does not depend on
$\omega$. In the Cartesian coordinate system with
$\hat{\mathbf{e}}_{z}$ along $\bm{B}$, the tensors of vacuum
polarizability $\bm{\chi}^\mathrm{vac}$ and permeability
become diagonal, with the coefficients determined by only three
numbers, called vacuum polarizability and magnetization coefficients.
These coefficients were obtained in \cite{Adler} at $b\ll1$ and in
\cite{HH97} at $b\gg1$. At arbitrary $b$, they were
calculated in \cite{KohriYamada} and fitted by simple analytic
functions in \cite{KK}. Convenient expressions of the components of
$\bm{e}_j$ through the components of $\bm{\chi}(\omega)$ and the
vacuum polarization and magnetization coefficients have been presented
in \cite{HoLai03}.

In the approximation of isotropic scattering, at a given
frequency $\omega$, the opacities can be written in the
form (e.g., \cite{KaminkerPS82})
\begin{equation}
   \opac_j^\mathrm{a} = \sum_{\alpha=-1}^1
     |e_{j,\alpha}(\theta_{kB})|^2 \,
        \frac{\sigma_\alpha^\mathrm{a}}{\mion},
\quad
 \opac_{jj'}^\mathrm{s} \!=\!\!
     {\frac34}
\!\!
  \sum_{\alpha=-1}^1 \!\!
     |e_{j,\alpha}(\theta_{kB})|^2 \,
     \frac{\sigma_\alpha^\mathrm{s}}{\mion}\int_0^\pi \!\!\!
       |e_{j',\alpha}(\theta_{kB}')|^2\sin\theta_{kB}'\,\mathrm{d}\theta_{kB}',
\label{opac}
\end{equation}
where $\alpha=0,\pm1$, and $\sigma_\alpha^\mathrm{a,s}$ are the
absorption and scattering cross-sections for the three basic cyclic
polarizations with $\bm{e}_0=\bm{e}_z$ and $\bm{e}_{\pm1} =
(\bm{e}_x\pm\mathrm{i}\bm{e}_y)/\sqrt{2}$. The
absorption cross-sections include contributions of photon interaction with
free electrons or ions in the plasma environment (free-free transitions) as
well as  with bound states of atoms and ions (bound-bound and bound-free
transitions).

At large optical depth 
$
   \tau_\omega =
\int_r^\infty \opac_\omega^\mathrm{eff}(r') \, \dd \ycol(r'),
$
where $\opac_\omega^\mathrm{eff}$ is the effective opacity defined
below, radiation is almost isotropic: the magnitude of the diffusive
radiative flux $F_{\omega,j}$
is much smaller than the mean intensity $J_{\omega,j}$, where
\beq
J_{\omega,j} =
\frac{1}{4\pi}\int_{(4\pi)}I_{\omega,j}(\khat)\,\dd\Omk{\bm{k}},
\qquad
    \bm{F}_{\omega,j}=\int_{(4\pi)}
         I_{\omega,j}(\khat)\,\khat\,\dd\Omk{\bm{k}}.
\eeq
In this case an approximate solution to (\ref{RTEmag}) is provided by
the diffusion approximation \cite{KaminkerPS82}:
\beq
I_{\omega,j}(\khat)\approx J_{\omega,j} +
\frac{3}{4\pi}\bm{F}_{\omega,j}\cdot\khat,
\quad
   \frac{\dd}{\dd\ycol}
D_{\omega,j}
         \frac{\dd}{\dd\ycol} J_{\omega,j} =
        \bar{\opac}_{\omega,j}^\mathrm{a}\,
       \left[ J_{\omega,j} -
           \frac{\mathcal{B}_{\omega,T}}{2} \right]
+
 \bar{\opac}_{\omega}^\mathrm{s}
        \left[J_{\omega,j} - J_{\omega,3-j} \right].
\label{diffmag}
\end{equation}
Here, the average absorption and scattering opacities are
\beq
   \bar{\opac}_{\omega,j}^\mathrm{a} = \frac{1}{4\pi}\int_{(4\pi)}
\opac_{\omega,12}^\mathrm{a}\,\dd\Omk{\bm{k}},
\qquad
\bar{\opac}_{\omega}^\mathrm{s} = \frac{1}{4\pi}
\int_{(4\pi)}\dd\Omk{\bm{k}'}
\int_{(4\pi)} \dd\Omk{\bm{k}}\,\,
\opac_{\omega,12}^\mathrm{s}(\khat',\khat) \, ,
\eeq
and the diffusion coefficient is
\beq
   D_{\omega,j} = 
\frac{\cos^2\theta_{nB}}{3\opac_{\omega,j}^\|} +
\frac{\sin^2\theta_{nB}}{3\opac_{\omega,j}^\perp},
\quad\mbox{where}\quad
\left\{
   \begin{array}{c}
 (\opac_{\omega,j}^\|)^{-1}
\\
   (\opac_{\omega,j}^{\perp})^{-1\rule{0pt}{2ex}}
   \end{array}
  \right\}
 = \frac34 \int_0^\pi
\left\{
   \begin{array}{c}
2\cos^2\theta_{kB} \\
\sin^2\theta_{kB}
\end{array}
  \right\}
\frac{\sin\theta_{kB}\,\mathrm{d}\theta_{kB}}{\opac_{\omega,j}(\theta_{kB})}\,.
\label{kappa-eff}
\eeq
The effective opacity for non-polarized
radiation is
$
\opac_\omega^\mathrm{eff}={2}/(3D_{\omega,1}+3D_{\omega,2}).
$
The diffusion approximation (\ref{diffmag}) serves as a
starting point to an iterative method \cite{SZ95}, which
allows one to solve the system (\ref{RTEmag})
more accurately.

To solve the radiative transfer equations, one must know the
dependencies of the temperature and densities of atoms, ions, and
electrons on the depth. These dependencies can be found from the
equations of thermal, hydrostatic, and ionization equilibrium
supplemented with the EOS. In the plane-parallel limit, the condition
of radiative thermal equilibrium requires the outward flux 
\beq
   F_\mathrm{ph} = \int_0^\infty\dd\omega 
      \int_{(4\pi)} \sum_{j=1}^2 I_{\omega,j}(\khat)
      \,\cos\theta_k\,\dd\Omk{\bm{k}}
\label{Fph}
\eeq
to remain constant throughout the atmosphere. In the approximation of
isotropic scattering, the equations of
hydrostatic balance are (cf.{} \cite{SuleimanovPouW12})
\beq
   \frac{\dd P}{\dd \ycol} = g - g_\mathrm{rad},
\qquad
   g_\mathrm{rad} 
\approx
    \frac{1}{c} 
   \int_0^\infty \dd\omega\,
    \int
    \cos\theta_k\, \sum_{j=1}^2
     \opac_{\omega,j}(\khat)\,I_{\omega,j}(\khat)
       \,\dd\Omk{\bm{k}}.
\hspace*{2em}\label{grad}
\eeq

\subsection{Physics input}
\label{sec:physinput}

To solve the radiation transfer problem formulated in
Sect.~\ref{sec:RT}, we need to know the composition of the plasma, its
EOS, and absorption and scattering cross-sections. All these
ingredients are affected by a strong magnetic field. There have been
many works devoted to these effects. We will give
only a brief sketch of the basic results (see \cite{P14} for details).

The motion of charged particles in a magnetic field $\bm{B}$ is
quantized in discrete Landau levels, whereas the longitudinal
(parallel to $\bm{B}$) momentum of the particle is continuous. The
threshold excitation energy of the $N$th Landau level equals $E_N=\mel
c^2 \,(\sqrt{1+2bN}-1)$. The wave functions that describe an electron
in a magnetic field \cite{SokTer} have a characteristic transverse
scale of the order of the ``magnetic length'' $\am=(\hbar
c/eB)^{1/2}=\aB/\sqrt{\gamma}$, where $\aB=\hbar^2/\mel e^2$ is the
Bohr radius.

In practice, Landau quantization becomes important when $N$ is not too
large. For the electrons in a plasma at equilibrium this means that
the electron cyclotron energy $\hbar\omc$ is at least comparable to
both the electron Fermi energy $\EF$ and temperature $\kB T$ in energy
units.  If $\hbar\omc$ is appreciably larger than both these energies,
then most electrons reside on the ground Landau level, and the field
is called \emph{strongly quantizing}. 
For a plasma of electrons and ions with charge $Ze$ and mass
$\mion=Am_\mathrm{u}$, where $m_\mathrm{u}=1.66054\times10^{-24}$~g,
the field is strongly quantizing if both conditions 
$\rho<\rho_B$ and $\zete\gg1$ are satisfied, where
\begin{equation}
  \rho_B =
 \frac{\mion}{\pi^2\sqrt2\,\am^3\,Z}
  = 7045 \,\frac{A}{Z}
       \,B_{12}^{3/2}\textrm{ \gcc},
\qquad
   \zete = \frac{\hbar\omc}{\kB T} = 134.34\,
   \frac{B_{12}}{T_6} .
\label{zeta_e}
\end{equation}
In neutron star atmospheres, as a rule,
the fields  $B\gg10^{10}$~G are strongly quantizing.
In the opposite case, where electrons are smoothly distributed over a
large number of the Landau levels, the field can be considered as
\emph{non-quantizing}. In the magnetospheres, which have lower
densities, electrons can condensate on the lowest Landau level even at
$B\sim10^8$~G because of the violation of the LTE conditions (e.g.,
\cite{Mesz}), but this is not the case in the photospheres (see
\cite{P14}).

For ions, the
parameter $\zete$ is replaced by 
$
   \zeti = \hbar\omci/\kB T = 0.0737\,(Z/A)
           B_{12}/T_6,
$
where $\omci=ZeB/(\mion c)$ is the ion cyclotron frequency.
In magnetar atmospheres, the parameter $\zeti$ is not small,
and the quantization of the ion motion should be
taken into account. A parameter analogous to $\rho_B$ is
unimportant for ions, because they are non-degenerate in
neutron star atmosphere.

Scattering cross-sections in neutron star atmospheres are well known
\cite{Ventura79,KaminkerPS82,Mesz}. The scattering on electrons is
strongly reduced compared to the zero-field case at $\omega\ll\omc$
and exhibits a cyclotron resonance at $\omc$. The ion scattering 
cross-section is analogous. In superstrong fields one cannot neglect the
scattering on ions, because of the resonance at $\omci$. In addition,
in  a quantizing magnetic field, a photon can be absorbed or emitted
by a free electron in a transition between Landau levels. In the
non-relativistic or dipole approximation, such transitions occur
between the neighboring levels at the fundamental cyclotron frequency
$\omc$. In the relativistic theory, the multipole expansion gives rise
to cyclotron harmonics \cite{Zheleznyakov}. They manifest themselves
in the opacities when one goes beyond the cold plasma approximation.
Absorption cross-sections at these harmonics were derived in
\cite{PavlovSY80} and represented in a convenient form in
\cite{SuleimanovPW12}. 

The quantization of electron motion leads to the appearance of
cyclotron harmonics in the free-free absorption as well, even in the
non-relativistic theory. Photon absorption cross-sections for an
electron, which moves in a quantizing magnetic field and interacts with
a non-moving point charge, were derived in \cite{PavlovPanov}. However,
an ion can be considered as non-moving only if $\omega \gg \omci$
\cite{Ginzburg}. In the
superstrong field of magnetars, the latter condition is unacceptable.
A more accurate treatment of absorption of a photon by the system of a
finite-mass ion and an electron was performed in \cite{PC03,P10}.
According to the latter studies, the free-free absorption cross-section
$\sigma_\alpha^\mathrm{a(ff)}(\omega)$ has peaks at the
multiples of both the electron and ion cyclotron frequencies for all
polarizations $\alpha$. However, these two types of peaks are
different. Unlike the electron cyclotron harmonics, the ion cyclotron
harmonics are so weak that they can be safely neglected in the neutron
star atmospheres.

As first noticed in \cite{CLR70}, atoms with bound states should be
much more abundant at $B\gg B_0$ than at $B\lesssim B_0$ in a neutron
star atmosphere at the same temperature. This difference is caused by
the increase of atomic binding energies and decrease of atomic sizes
at $B\gg B_0$. Therefore it is important to consider the bound states
and bound-bound and bound-free transitions in a strong magnetic field
even for light-element atmospheres, which would be almost fully
ionized in the non-magnetic case.

Many authors studied atoms with an infinitely heavy (fixed in space)
nucleus in strong magnetic fields (see, e.g., \cite{Ruder_ea,Lai01}, for
review). This model, however, can be considered only as a
first approximation. If $\gamma=B/B_0$ is not negligibly small compared to
the nucleus-to-electron mass ratio $\mion/\mel$, one should take into
account quantum oscillations of an atomic nucleus, which are different
for different atomic quantum states. Moreover, the astrophysical
simulations assume finite temperatures, hence thermal motion of
particles. The theory of motion of a system of point charges $q_i$ in
a constant magnetic field was reviewed in \cite{JHY83}. Instead of the
canonical momentum $\bm{P}$, a relevant conserved quantity is
pseudomomentum 
$
   \bm{K}=\bm{P}+(1/2c)\,\bm{B}\times\sum_i q_i \bm{r_i}.
$
For a hydrogen atom,
$
   \bm{K} = \bm{P} - ({e}/{2c})\,\bm{B}\times \bm{r},
$
where $\bm{r}$ connects the proton and the electron
\cite{GorkovDzyal}. The specific effects related to collective motion
of a system of charged particles are especially important in neutron
star atmospheres at $B\gg B_0$. In particular, so called
\emph{decentered states} may become populated, where an electron is
localized mostly in a ``magnetic well'' aside from the Coulomb center.
Numerical calculations of the energy spectrum of the hydrogen atom
with allowance for the effects of motion across a strong magnetic
field were performed in  \cite{VDB92,P94}. Probabilities of various
radiative transitions for a hydrogen atom moving in a strong magnetic
field were studied in \cite{P94,PP95,PP97}.

Quantum-mechanical calculations of the characteristics of a He$^+$ ion
moving across a strong magnetic field are performed in
\cite{BezchastnovPV98,PB05}. The basic difference from the case of a neutral
atom is that the the ion motion is restricted by the field in the
transverse plane, therefore the values of $K^2$ are quantized (see
\cite{JHY83}).

Currently there is no detailed calculation of binding energies,
oscillator strengths, and photoionization cross-sections for atoms and
ions other than H and He$^+$, arbitrarily moving in a strong magnetic
field. For such species one usually neglects the decentered states and
uses a perturbation theory with respect to $\Kp$ \cite{VB88,PM93}. An
order-of-magnitude estimate \cite{P14} of the validity of the latter
gives, for an atom with mass $m_\mathrm{a}=Am_\mathrm{u}$, the
condition $\kB T/E_\mathrm{b}\ll m_\mathrm{a}/(\gamma
\mel)\approx4A/B_{12}$, where $E_\mathrm{b}$ is the atomic ionization
energy. It is satisfied for low-lying levels of carbon and heavier
ions, if $B\lesssim10^{13}$~G and $T\lesssim10^6$~K (see the
discussion in \cite{P14}). A practical method of calculations of the
quantum-mechanical characteristics of multielectron atoms and ions,
based on a combination of several perturbation theories with respect
to different physical parameters, has been developed in
\cite{MoriHailey02}.

Since the quantum-mechanical characteristics of an atom in a strong
magnetic field depend on $K_\perp$, the atomic distribution over
$K_\perp$ cannot be written in a closed form, and only the
distribution over longitudinal momenta $K_z$ remains Maxwellian. The
first EOS calculations with the complete account of these effects have
been performed in \cite{PCS99} for hydrogen atmospheres. The same
approach with slight modifications was then applied to strongly
magnetized helium plasmas \cite{MoriHeyl}.

The calculation of the cross-section of photoabsorption by
bound states of atoms and ions in the strong magnetic fields implies
averaging of the cross-sections over all
values of $K_\perp$ at every $\omega$. Since the distribution over
$K_\perp$ is continuous for the atoms and  discrete for the ions, such
averaging implies an integration over $K_\perp$ for atoms and a
summation for ions. Statistical weights for this averaging should be
consistent with the statistical model of the plasma used in the EOS
calculations. To date, such fully self-consistent calculations,
including both centered and decentered bound states (i.e., small and
large $K_\perp$) have been realized only for neutron-star atmospheres
composed of hydrogen \cite{PC03,PC04,PCH14}. For atoms and ions with
several bound electrons (C, O, Ne), calculations have been performed
\cite{MoriHailey06,MoriHo} in terms of the above-mentioned
perturbation theory.

\subsection{Modeling observed spectra}
\label{sec:models}

The strong gravity of a neutron star induces a significant redshift of
the local photon frequency $\omega$ to  $\omega_\infty =
\omega/(1+z_g)$ in the remote observer's reference frame. Accordingly,
a thermal spectrum with effective temperature $\Teff$ transforms for
the remote observer into a spectrum with a lower ``observed''
temperature $\Teff^\infty = \Teff / (1+z_g)$.  Along with the radius
$R$ that is determined by the equatorial length $2\pi R$ in the local
reference frame, one often considers an \emph{apparent radius} for a
remote observer, $ R_\infty = R \,(1+z_g)$, so that the apparent
photon luminosity $L_\mathrm{ph}^\infty$ is determined by the
Stefan-Boltzmann law $ L_\mathrm{ph}^\infty = L_\mathrm{ph}/(1+z_g)^2
=4\pi\sSB\,R_\infty^2 \,(\Teff^\infty)^4, $ where $\sSB$ is the
Stefan-Boltzmann constant, and $L_\mathrm{ph} = 4\pi \sSB\,R^2
\Teff^4$ is the luminosity in the local reference frame.

Since thermal diffusion is anisotropic in a strong magnetic field, the
temperature is non-uniform over the stellar surface, which can lead to
modulation of the observed thermal flux as the star rotates
\cite{GreensteinHartke}. Then it is convenient to define a local
effective surface temperature $\Ts$ by the relation $
F_\mathrm{ph}(\theta,\varphi) = \sSB\Ts^4, $ where $F_\mathrm{ph}$
is the local radial flux density at the surface point, \req{Fph},
determined by the polar angle ($\theta$) and azimuth ($\varphi$) in
the spherical coordinate system. Since $L_\mathrm{ph}$ is the integral
of $F_\mathrm{ph}(\theta,\varphi)$ over the surface, $\Teff^4$ is
equal to the average of $\Ts^4$.

Because of the light bending in strong gravity, the observer receives
a photon whose wave vector makes a different angle $\theta$ with
normal to the surface than the angle $\theta_k$ at the point where
this photon was emitted. For the compactness parameters $x_g<1/2$,
this effect is well described by the simple approximation
\cite{Beloborodov02} $ \cos\theta_k = x_g + (1-x_g)\cos\theta$. The
spectral flux that comes to an observer can be most easily calculated
using the equations presented in \cite{PoutanenBeloborodov06} provided
that the distribution of $I_{\omega,j}(\khat)$ is known for the entire
visible surface of the neutron star. The problem is complicated
because of non-uniform surface distributions of the magnetic field and
effective temperature. A fiducial model for the magnetic field
distribution is the relativistic dipole \cite{GinzburgOzernoi}, but
recent numerical simulations of the magnetothermal evolution produce
more complicated distributions (see \cite{Vigano_ea13,Elfritz_ea16}
and references therein). The temperature distribution, consistent with
the magnetic-field distribution, is found from calculations of heat
transport in neutron star envelopes (see \cite{PPP15} for review).

Local spectra from strongly magnetized, fully ionized neutron-star
atmospheres  were calculated in \cite{Shibanov_ea92,Zavlin_ea95}.
Later the authors
included spectral tables for $B=10^{12}$ and $10^{13}$~G
in the \textit{XSPEC} package \cite{XSPEC} under the name \textsc{nsa}.
 The authors have shown that the spectra of magnetic
hydrogen and helium atmospheres are softer than the respective
non-magnetic spectra, but harder than the blackbody spectrum with the
same temperature. At contrast to
the isotropic blackbody radiation, radiation of a magnetic atmosphere
consists of a narrow ($<5^\circ$) pencil beam along the magnetic field
and a broad fan beam with typical angles  $\sim20^\circ-60^\circ$ (as
had been already predicted in \cite{GnedinSunyaev74}).

Later, similar calculations were performed by other research groups
\cite{Zane_ea01,HoLai03,vanAdelsbergLai}, paying special attention
to possible manifestations of the ion cyclotron resonance in magnetar
spectra. In particular, it was shown
\cite{LaiHo02,HoLai03,vanAdelsbergLai}
that vacuum polarization leads to conversion of the normal modes: a
photon related to one mode transforms, with a certain probability, into
a photon of the other mode while crossing a surface at a
critical density which depends on $\omega$. For $B\sim10^{14}$~G and
$\omega\sim\omci$, conversion occurs at a density where the atmosphere is
optically thin for the extraordinary mode but optically thick for the
ordinary mode. As a result, ordinary photons converting into extraordinary
ones will reduce the strength of the ion cyclotron absorption feature in the
emission spectrum of magnetars.

A strongly magnetized hydrogen atmosphere model with full account of
the partial ionization and the atomic motion effects was constructed
in \cite{KK}. The calculated spectra revealed a narrow absorption line
at the proton cyclotron energy and some features related to atomic
transitions. Similar to fully ionized plasma models, the
intensity has a maximum at higher energies relative to the maximum of
the Planck function, but at lower energies relative to the non-magnetic
hydrogen atmosphere model. Therefore, the model of a fully-ionized
atmosphere with a strong magnetic field can yield a relatively reliable
temperature, although it does not reproduce the spectral features
caused by atomic transitions.

\begin{figure}[t]
  \includegraphics[width=.49\linewidth]{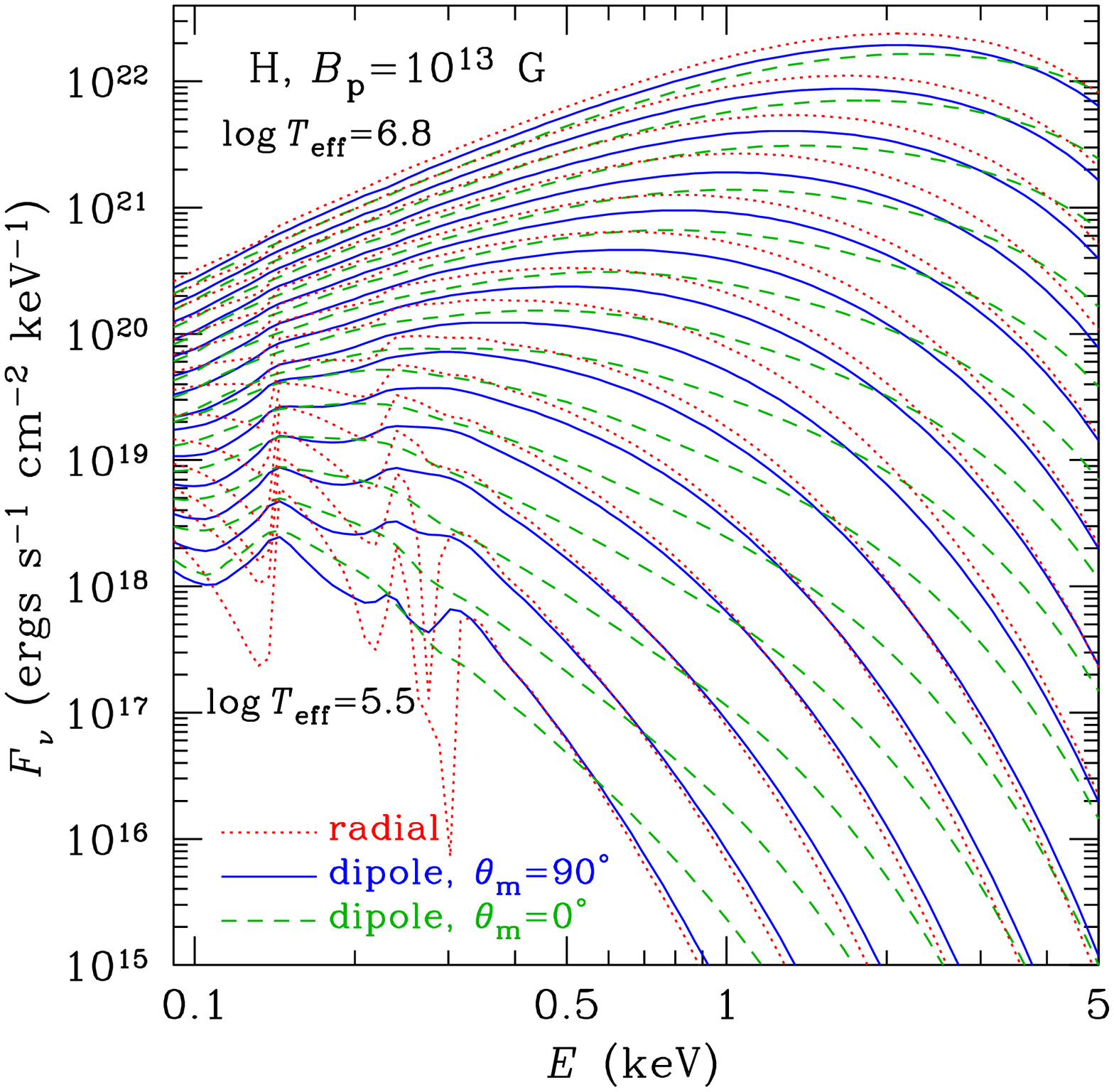}
  \includegraphics[width=.505\linewidth]{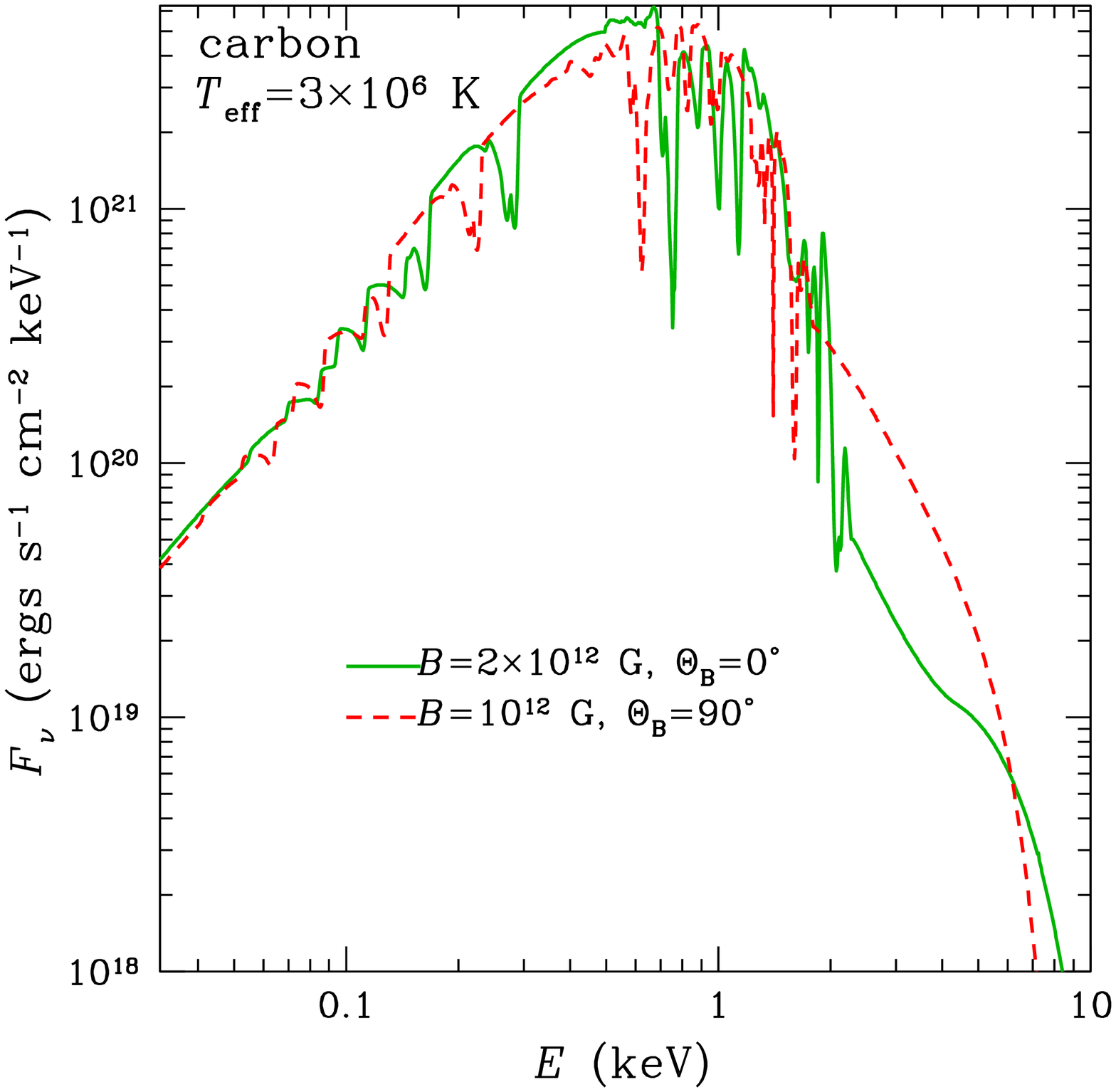}
\caption{\emph{Left panel.} Integral spectra of a hydrogen atmosphere
of a neutron star with $M=1.4\,M_\odot$, $R=12$ km, and with different
effective temperatures $\Teff$ ($\log\Teff$ (K) from 5.5 to 6.8 with
step 0.1). The dashed and solid lines represent the model with a
dipole field of strength $B_\mathrm{p}=10^{13}$~G at the pole and
oriented along and across the line of sight, respectively. For
comparison, the dotted curves show the model with a constant field
$B=10^{13}$~G, normal to the surface.
\emph{Right panel (from \cite{MoriHo}).} Local spectra at the magnetic
pole (solid curve) and equator (dashed curve) for a neutron star with
carbon atmosphere, the dipole field with polar strength of
$B_\mathrm{p}=2\times10^{12}$~G (neglecting the relativistic
corrections) and uniform effective temperature $3\times10^6$~K.
}
\label{fig:spectr13}
\end{figure}

Synthetic spectra of partially ionized hydrogen atmospheres with
averaging over the
stellar surface with realistic temperature distribution were
calculated in \cite{HoPC}. The left panel of Fig.~\ref{fig:spectr13}
shows examples of such spectra calculated for a neutron star
with $M=1.4\,M_\odot$ and $R=12$~km, endowed with a dipole magnetic
field, whose value at the magnetic pole is $B_\mathrm{p}=10^{13}$~G.
For comparison, the dotted lines show spectra calculated for constant
$B=10^{13}$~G for the same values of $\Teff$. In the dipole
configuration, the spectral features are strongly smeared by the
averaging over the surface temperature and magnetic field distributions,
and the spectrum depends on the magnetic
axis orientation $\theta_\mathrm{m}$ relative to the line of sight.
As the star rotates, the latter dependence leads to a rotational phase
dependence of the spectra. 

Models of partially ionized neutron-star atmospheres composed of
strongly ionized iron, oxygen, and neon were constructed in
\cite{RRM97,MoriHailey06,MoriHo}. The effects related to the finite
nuclear masses were treated in the first order of the perturbation
theory (see Sect.~\ref{sec:physinput}).
The right panel of Fig.~\ref{fig:spectr13} shows local spectra of a
strongly ionized carbon atmosphere at the magnetic pole and equator of
a neutron star for the non-relativistic magnetic dipole model with
$B_\mathrm{p}=2\times10^{12}$~G. It is clear that in this case the
averaging over the surface, with spectra varying between these two
extremes, would also result in smearing of some of the features seen
in the local spectra.

The calculated spectra of partially ionized, strongly magnetized
neutron star atmospheres composed of hydrogen, carbon, oxygen, and
neon with magnetic fields $B\sim10^{10}-10^{13}$~G
(see Sect.~\ref{sec:moderate}) are included
in \textit{XSPEC} \cite{XSPEC} under the names \textsc{nsmax}
\cite{MoriHo,HoPC} and \textsc{nsmaxg} \cite{MoriHo,Ho14,PCH14},
with the latter allowing for varying surface gravity.\footnote{We thank Peter
Shternin for finding an error in the old (prior to 2014) version of
\textsc{nsmax}.}

\subsection{Moderately strong magnetic fields}
\label{sec:moderate}

Most of the thermally emitting INSs have surface magnetic fields in
the range $10^{12}\mbox{ G}\lesssim B \lesssim 10^{15}$~G, but one
class of sources, so called central compact objects (CCOs) have
probably surface fields $B\lesssim10^{11}$~G
\cite{HalpernGotthelf10,GotthelfHA13,Ho13}.
These fields are strong enough to radically affect properties of
hydrogen atoms and strongly quantize the electrons in the neutron-star
atmosphere, but they are below the field strengths previously
available in the atmosphere model \textsc{nsmaxg} \cite{Ho14}. The
possibility to extend the model to the field strengths
$3\times10^{10}\mbox{ G}\lesssim B\lesssim 10^{12}$~G, which we call
\emph{moderately strong}, has opened recently. The task of calculation
of such models is arduous because the dimensionless field parameter
$\gamma=B/B_0$ is smaller than in the previously calculated models,
which entails the need to include more terms than previously in the
wave-function expansion over the set of Landau orbitals \cite{P94}. In
addition, with decreasing $B$, the energy spectrum of the bound states
of a magnetized atom becomes denser, which necessitates inclusion of
more such states in the consideration. Meanwhile, since $\gamma\gg1$,
the center-of mass motion of the atom noticeably affects the atomic
properties. 

In order to solve this problem, we constructed analytical fitting
formulae for atomic energies, sizes, and main oscillator strengths as
functions of $B$, discrete quantum numbers of initial and final
states, and  pseudomomentum $\Kp$ \cite{PCH14}. For bound-free
transitions, we calculated extensive tables of cross-sections as
functions of $\Kp$ and photon frequency $\omega$ for a number of bound
states at every given $B$ and interpolated across these tables. This
method allowed us to calculate the opacities for moderately magnetized
hydrogen atmospheres \cite{PCH14}. In addition to the bound-bound,
bound-free, and free-free absorption, we also included the
cyclotron absorption with the electron cyclotron harmonics calculated
beyond the cold plasma approximation (Sect.~\ref{sec:physinput}).

\begin{figure}[t]
\begin{center}
\includegraphics[height=.49\linewidth]{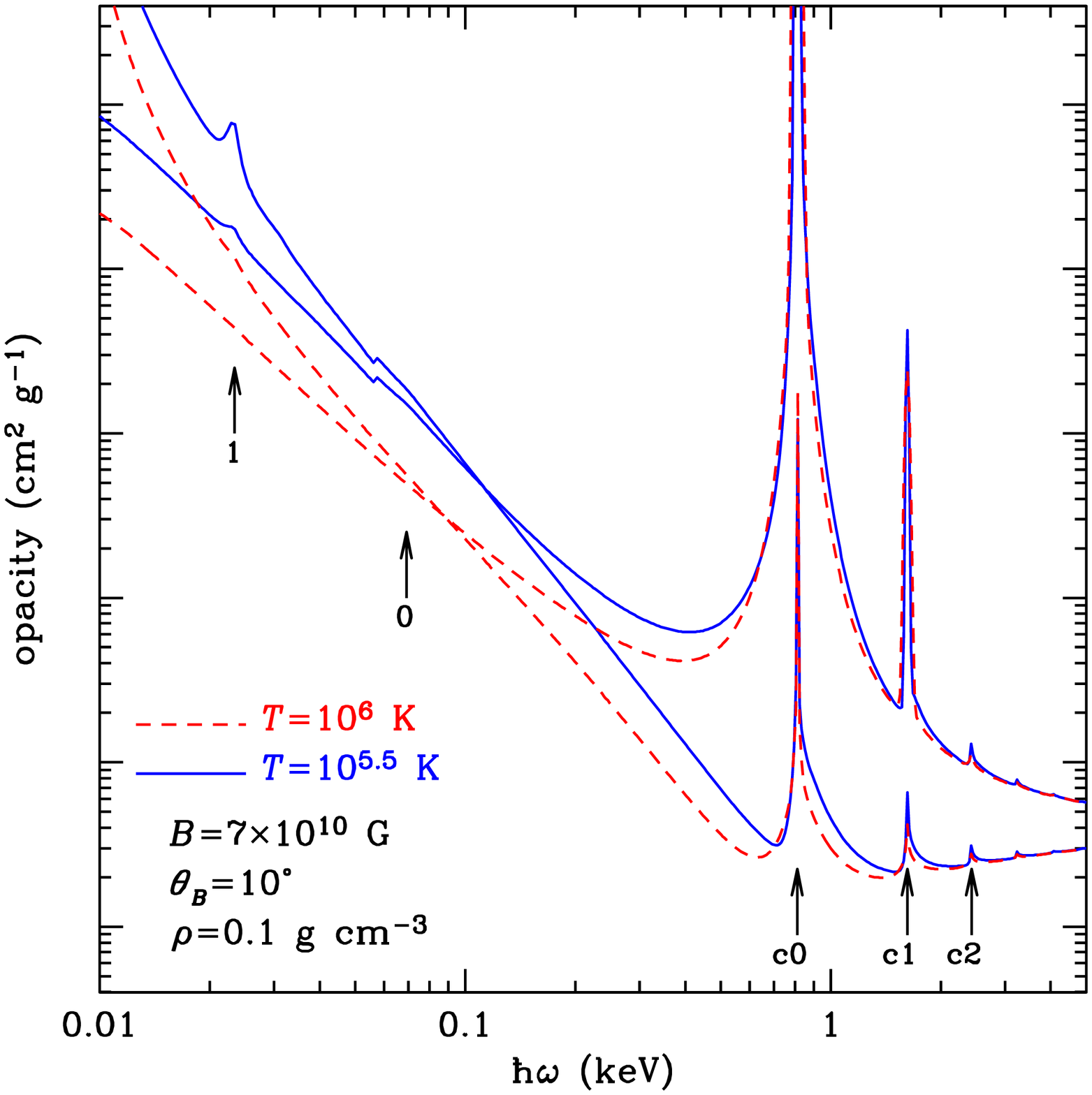}
\includegraphics[height=.49\linewidth]{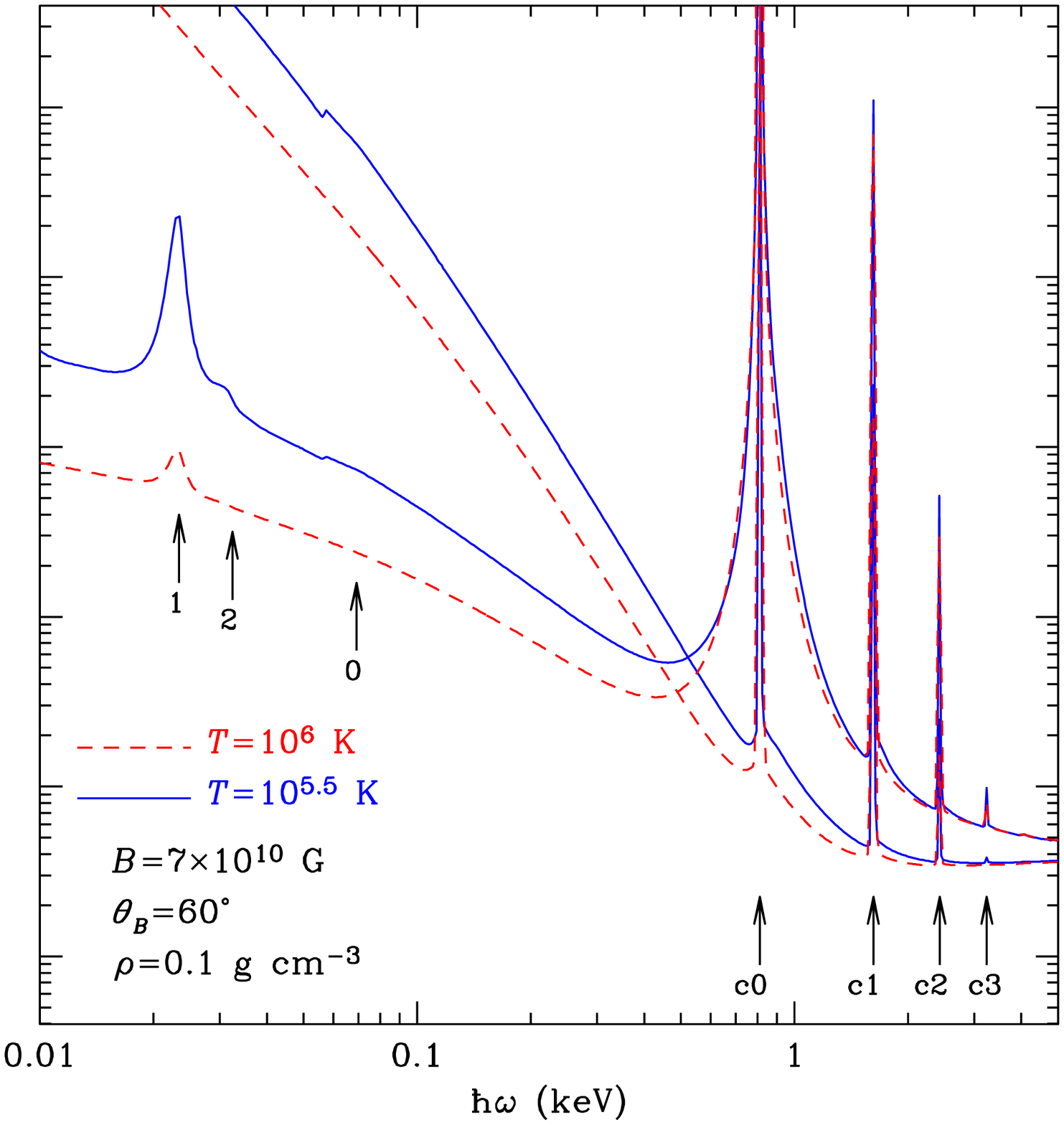}
\caption{Opacities of the normal polarization modes $j=1,2$
(the lower and upper curve of each type) at
$\rho=0.1$ \gcc{} and $B=7\times10^{10}$~G for
$\theta_{kB}=10^\circ$ (left panel), 
$\theta_{kB}=60^\circ$ (right panel),
$T=10^{5.5}$~K (solid curves, and $T=10^{6}$~K (dashed curves).
The arrows marked 0, 1, 2 correspond to different characteristic
transition energies of the \emph{non-moving} hydrogen atom: 0 -- ground state
binding energy, 1 -- transition between the ground state and first
excited state, 2 -- transition between the first and second excited
states. The arrows
marked c0, c1, c2, c3 correspond to cyclotron harmonics energies
$(N+1)\hbar\omc$ with $N=0,1,2,3$, respectively.
\label{fig:opac}
}
\end{center}
\end{figure}

\begin{figure}[t]
\begin{center}
\includegraphics[height=.49\linewidth]{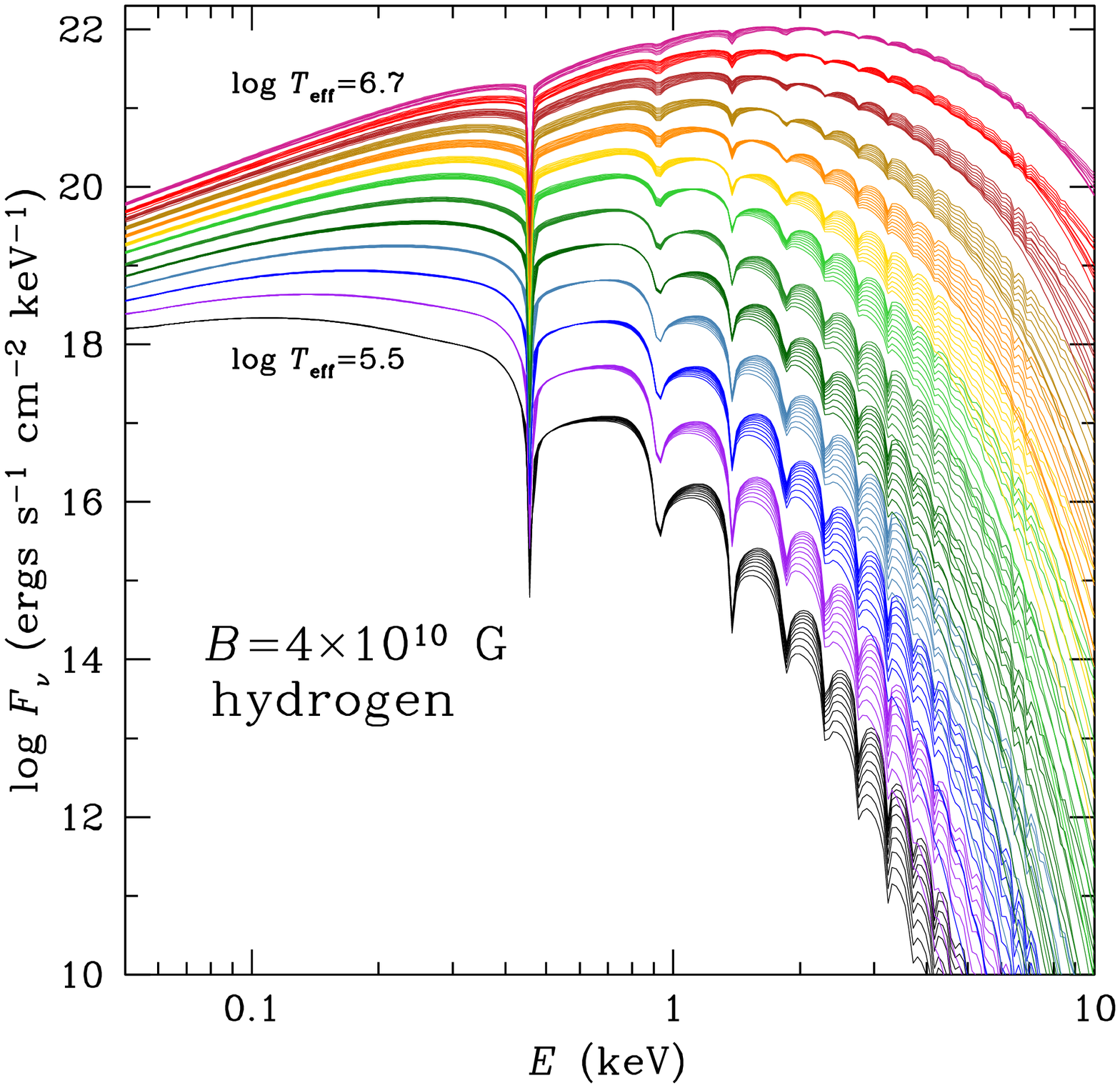}
\includegraphics[height=.49\linewidth]{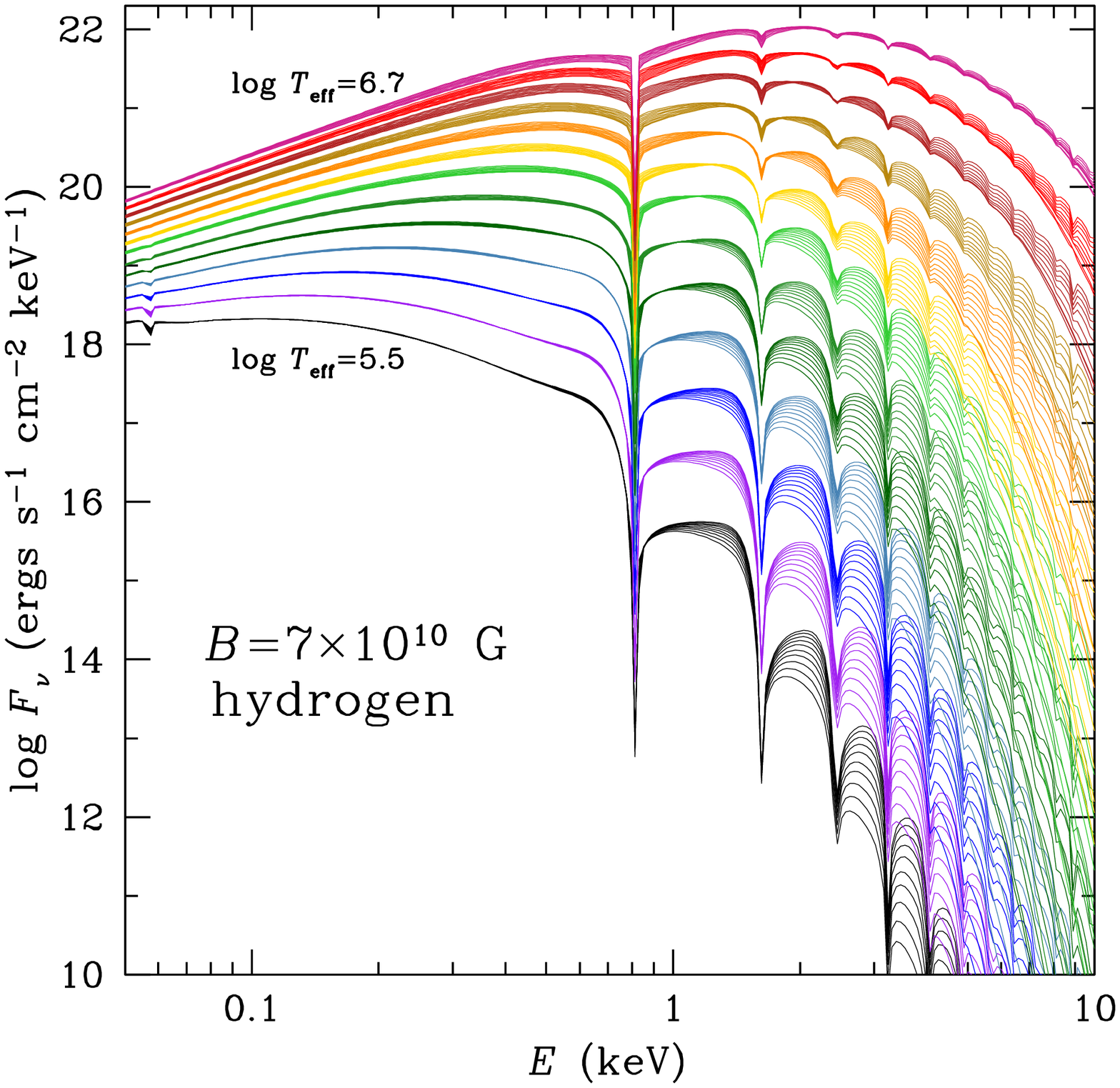}
\caption{Spectra for partially ionized hydrogen atmosphere models with
$B = 4\times10^{10}$~G (left panel) and $7\times10^{10}$~G (right
panel), with varying effective temperature $\Teff$ and surface gravity.
\label{fig:spectra}
}
\end{center}
\end{figure}

Examples of the calculated opacities are shown in Fig.~\ref{fig:opac}
for two values of $\theta_{kB}$ and two values of $T$. Atomic features
(marked 0, 1, 2) are smeared and shifted leftward due to the magnetic
broadening caused by the dependence of $E_\mathrm{b}$ on $\Kp$, 
which is much larger than
the usual Doppler broadening. At higher $T$, these features disappear
because of the thermal ionization. At contrast, the electron cyclotron
features (marked c0, c1, c2, c3) remain clearly visible even at high
$T$. The calculated opacities have been archived at the Strasbourg
astronomical Data
Center.\footnote{\texttt{http://www.aanda.org/articles/aa/abs/2014/12/aa24619-14/aa24619-14.html}}

Work on the extension of the database  \textsc{nsmaxg} to the
moderately magnetized atmosphere models, applicable to CCOs, is in
progress. Examples of the calculated spectra are shown in
Fig.~\ref{fig:spectra}.

\section{Condensed surfaces and thin atmospheres}
\label{sec:cond}

Ruderman \cite{Ruderman71} suggested that a strong magnetic field can
stabilize polymer chains directed along the field lines, and that the
dipole-dipole attraction of these chains may result in a condensation.
Later this conjecture was confirmed, although the binding and
sublimation energies proved to be smaller than Ruderman assumed (see
\cite{MedinLai06b} and references therein).

From the thermodynamics point of view, the magnetic condensation is a
phase transition caused by the strong electrostatic attraction between
the ions and electrons in a dense plasma. In the absence of magnetic
field, such phase transitions were studied theoretically since 1930s
(see \cite{PPT} for a review). The simplest estimate of the phase
transition domain is obtained in the ion-sphere model
\cite{Salpeter61}, where the electrons are replaced by a uniform
negative background, and the potential energy per ion is estimated as
the electrostatic energy of the ionic interaction with the negative
background contained in the sphere of radius
$\aion=(4\pi\nion/3)^{-1/3}$, where $\nion=\nel/Z$ is the ion number
density. At sufficiently high density, the electrostatic
pressure is counterbalanced by the pressure of degenerate electrons.
By equating the sum of the (negative) electrostatic pressure and the
kinetic pressure of the electron gas to zero, one obtains density
$\rhos$ of the condensed surface.

A strongly quantizing magnetic field lowers the electron Fermi
temperature, therefore the above-mentioned pressure balance is shifted
to higher densities. With increasing $\rhos$, the electrostatic
attraction becomes stronger and, consequently, the critical
temperature $\Tc$ of the phase transition also increases. For this
reason, the phase transition may be expected in the neutron star
envelopes  despite their relatively high temperatures
$T\sim(10^5-10^7)$~K. Lai \cite{Lai01} estimated the condensed-surface
density as  $ \rhos\approx 561\,\eta\,A Z^{-3/5}
B_{12}^{6/5}\mbox{~\gcc}, $ where $\eta$ is an unknown factor, which
would be equal to 1 in the ion sphere model. In more sophisticated
models $\eta$ can differ from 1 because of ion correlations,
electron-gas polarizability, and bound state formation. In particular,
the values of $\rhos$ calculated by Medin \& Lai
\cite{MedinLai06b,MedinLai07} can be reproduced with
$\eta=0.517+0.24/B_{12}^{1/5}\pm0.011$ for carbon and
$\eta=0.55\pm0.11$ for iron. The critical temperature $\Tc$, obtained
numerically in \cite{MedinLai06b,MedinLai07}, is approximately (within
a factor of 1.5) given by the expression $\Tc \approx
5\times10^4\,Z^{1/4}\,B_{12}^{3/4}$~K in the  magnetic field range
$10^{12}$~G $\lesssim B \lesssim 10^{15}$~G.

The thermal radiation of the condensed surface is determined by its
emissivities in two normal modes, which are related to the
reflectivities through the Kirchhoff law. Progressively improved
calculations of the surface emissivity were presented in
\cite{Itoh75,LenzenTruemper,Brinkmann80,TurollaZD04,PerezAMP05,surfem,reflefit}.
While the early works assumed that the ions are firmly fixed in the
crystalline lattice, more recent works
\cite{surfem,PerezAMP05,reflefit} consider also the alternative model
of free ions. The reality lies probably between these two extreme
limits, but this problem has not yet been definitely solved.

\begin{figure}[t]
\begin{center}
\includegraphics[height=.45\linewidth]{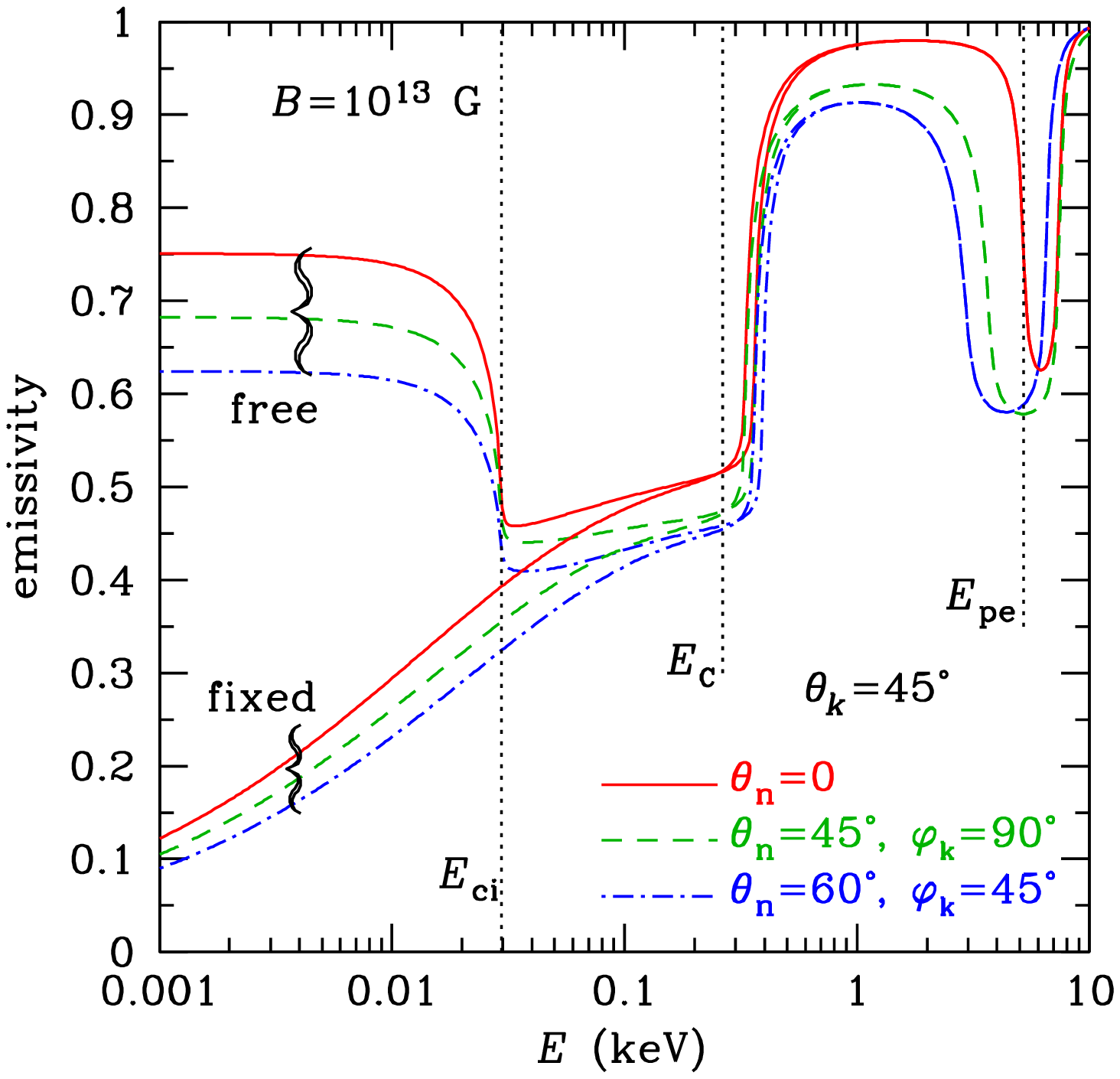}
\includegraphics[height=.45\linewidth]{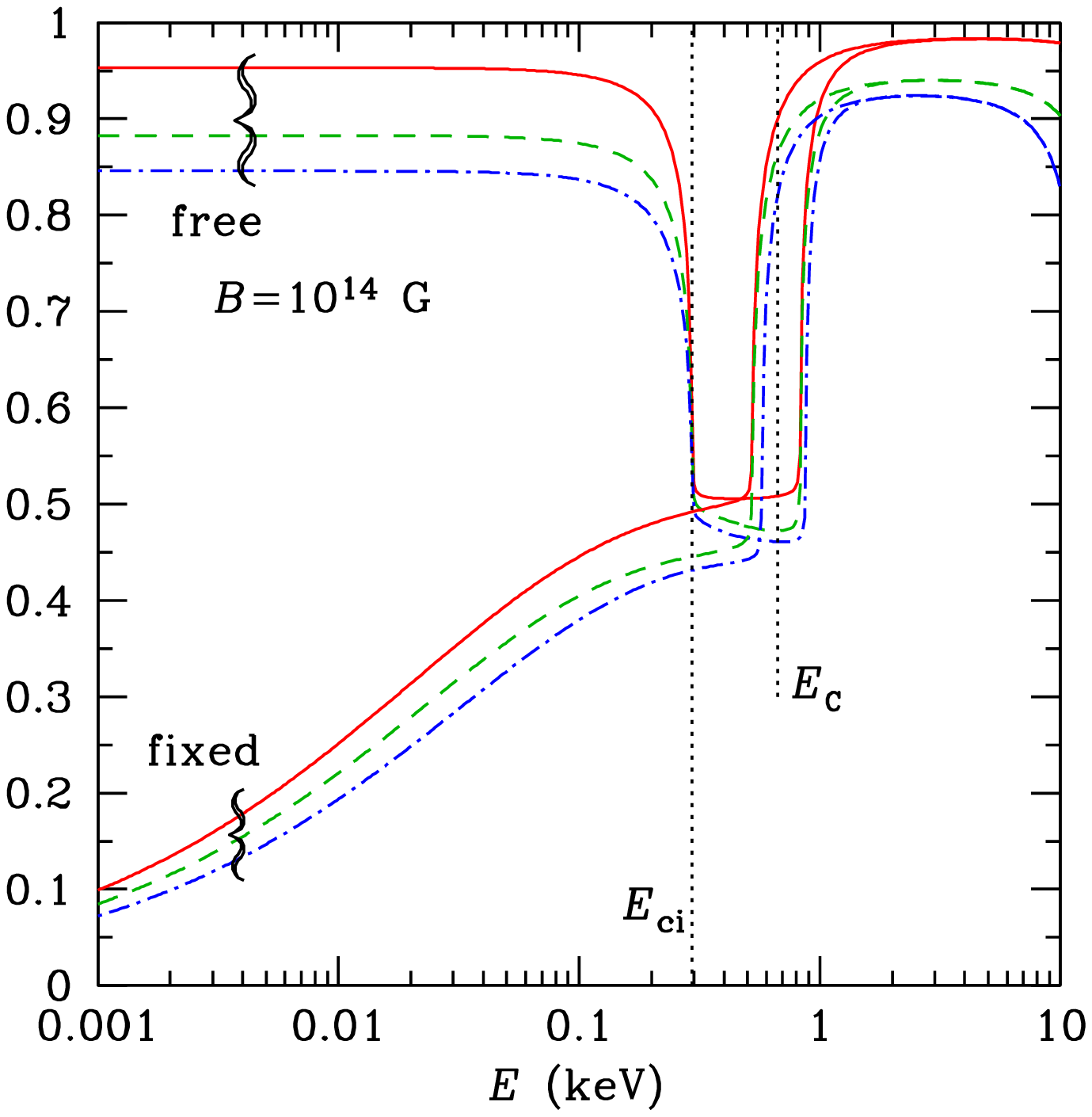}
\caption{Dimensionless emissivity of a condensed iron surface at
$B=10^{13}$~G (\emph{left panel}) and $10^{14}$~G (\emph{right
panel}), averaged over polarizations, is shown as a function of energy
of a photon emitted at the angle $\theta_k=45^\circ$, for different
angles $\theta_{nB}$ between magnetic field $\bm{B}$ and the
normal to the surface and angles $\varphi_k$ between the projections
of $\bm{B}$ and wave vector $\bm{k}$ on the surface. For each
parameter set,
curves marked ``free'' and ``fixed'' correspond to models with free
and fixed ions (see text).
Vertical dotted lines mark positions of the characteristic
energies: the ion cyclotron energy $E_\mathrm{ci}=\hbar\omci$, the
electron plasma energy $E_\mathrm{pe}=\hbar\ompe$, and the hybrid
energy $E_\mathrm{C}$.
\label{fig:rb}}
\end{center}
\end{figure}

The intensity of radiation from a condensed surface is equal to the
Planck function $\mathcal{B}_{\omega,T}$ [\req{Planck}] multiplied by
the normalized emissivity. Figure~\ref{fig:rb} shows examples of the
normalized emissivity as a function of photon energy $E=\hbar\omega$,
in both the free- and fixed-ion limits, for the wave-vector
inclination angle $\theta_k=45^\circ$, for $B=10^{13}$~G and
$10^{14}$~G, and different values of the magnetic-field inclination
$\theta_{nB}$ and azimuthal angles $\varphi_k$. The emissivity
rapidly changes near the ion cyclotron energy
$E_\mathrm{ci}=\hbar\omci$, the electron plasma energy
$E_\mathrm{pe}=\hbar\ompe$, and the energy
$E_\mathrm{C}=E_\mathrm{ci}+E_\mathrm{pe}^2/\hbar\omc$ (see
\cite{surfem} for explanation of these features).

Motch et al.~\cite{MotchZH03} suggested that some neutron stars can possess a
hydrogen atmosphere of a finite thickness above the solid iron
surface. If the optical depth of such atmosphere is small for some
wavelengths and large for other ones, the thermal spectrum is
different from that of thick atmospheres. Such spectra were calculated
in \cite{Ho_ea07,SuleimanovPW09,Suleimanov_ea10} using simplified
boundary conditions for the radiative transfer equation at the inner
boundary of the atmosphere. More accurate boundary conditions
\cite{reflefit} take into account that an extraordinary or ordinary
wave, falling from the atmosphere on the surface, gives rise to
reflected waves of both normal polarizations, whose intensities add to
the respective intensities of the waves emitted by the condensed
surface.

In general, local spectra of radiation emitted by thin hydrogen
atmospheres over a condensed surface reveal a narrow absorption line
corresponding to the proton cyclotron resonance in the atmosphere,
features related to atomic transitions broadened by the motion effects
(Sect.~\ref{sec:physinput}), and in the free-ions model also
a kink corresponding to the ion
cyclotron energy of the substrate ions.  Some of these features may be
absent, depending on the atmosphere thickness and magnetic field
strength. At high energies, the spectrum is determined by the
condensed-surface emission, which is softer than the spectrum of the
thick hydrogen atmosphere.

One may also consider an atmosphere having a helium layer
beneath the hydrogen layer. The spectrum of such ``sandwich
atmosphere'' can have two or three absorption
lines in the range $E\sim(0.2$\,--\,1) keV at $B\sim10^{14}$~G
\cite{SuleimanovPW09}.

\section{Polarization}
\label{sec:pol}

Thermal radiation emergent from neutron stars with strong magnetic
fields is expected to be strongly polarized. Since the opacity is
smaller for the extraordinary mode, with electric vector mainly
transverse to the magnetic field, this mode escapes from deeper and
hotter layers in the atmosphere, so that thermal radiation acquires
polarization perpendicular to the local magnetic field
\cite{PavlovShibanov78}. Polarization of the observed radiation
depends on the distribution of magnetic field and temperature over the
visible neutron star surface. As the star rotates, the polarization
pattern shows periodic variations, so that measuring the polarization
pulse profile allows one to constrain the orientation of the rotation
axis and the field strength and geometry \cite{PavlovZavlin00,LaiHo03a}.
Therefore, future X-ray polarization measurements are expected to
resolve degeneracies that currently hamper the determination of
magnetar physical parameters using thermal models
\cite{vanAdelsbergPerna09,Taverna_ea14}. 

Polarization properties of the radiation can be described by the
Stokes parameters $(I,Q,U,V)$, where $I$ is the total intensity, $Q$
and $U$  describe linear polarization, and $V$ circular polarization
\cite{ChandraRT}. In terms of intensities $I_\mathrm{O}$ and
$I_\mathrm{X}$ of the ordinary and extraordinary modes,
$I=I_\mathrm{X}+I_\mathrm{O}$,
$Q=(I_\mathrm{O}-I_\mathrm{X})p_\mathrm{lin}\cos2\phi$,
$U=(I_\mathrm{O}-I_\mathrm{X})p_\mathrm{lin}\sin2\phi$, where
$p_\mathrm{lin}$ is the degree of linear polarization of each normal
mode and $\phi$ is the angle between the major axis of the ordinary
mode and the $x$-axis of the reference frame in which the Stokes
parameters are defined \cite{PavlovZavlin00,LaiHo03b}.

The thermal emission of a condensed surface is also polarized, because
the reflectivities of normal modes are different. The polarization
depends on the photon energy $E=\hbar\omega$ and angles $\theta_{nB}$,
$\theta_k$, and $\varphi_k$ (Sect.~\ref{sec:cond}). In a local
spectrum, the degree of  polarization can reach tens percent in some
energy ranges. However, if radiation comes from a large surface area
with varying magnetic field and temperature, the net polarization
becomes much smaller.

\begin{figure}[t]
\begin{center}
\includegraphics[height=.45\linewidth]{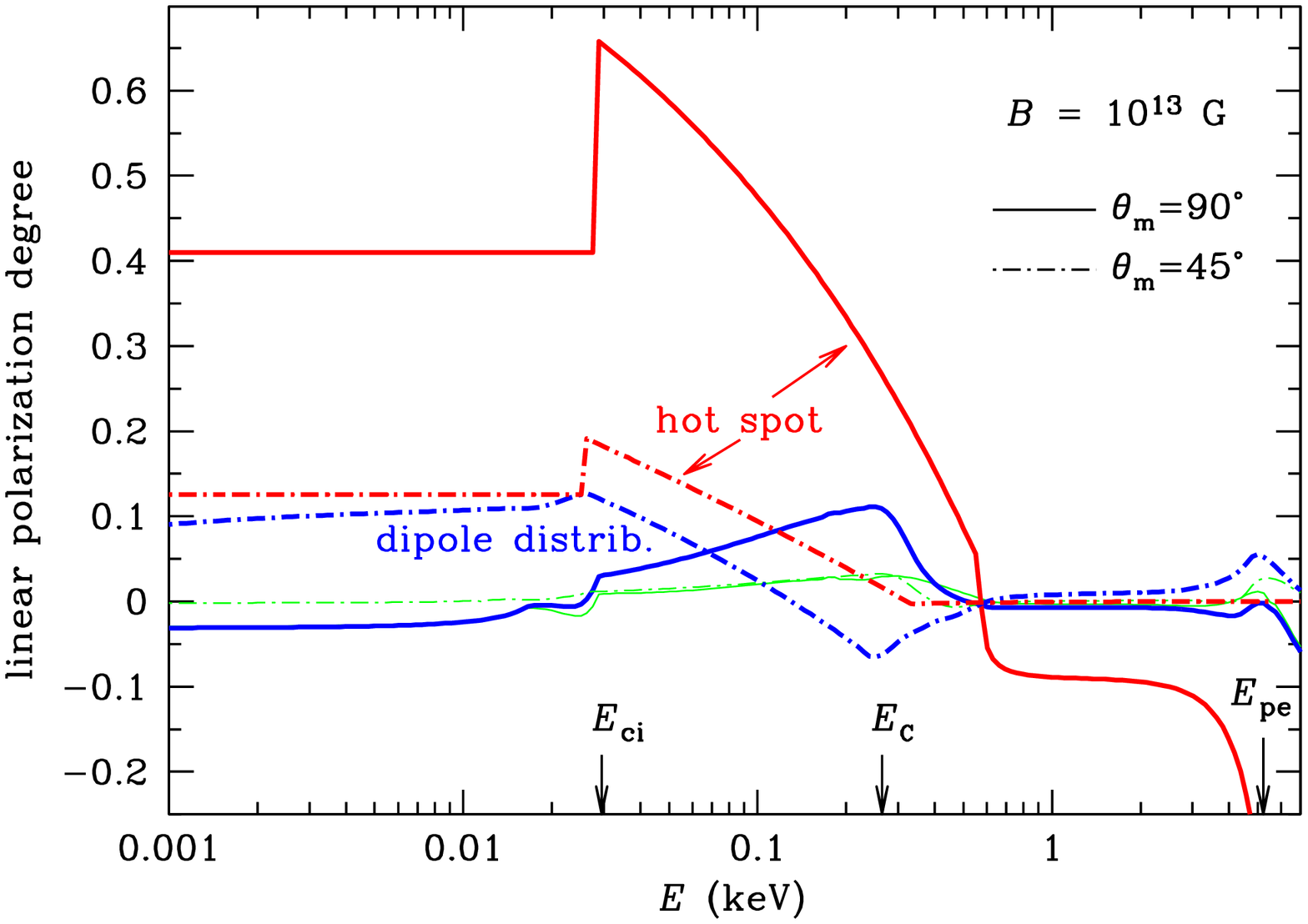}
\includegraphics[height=.45\linewidth]{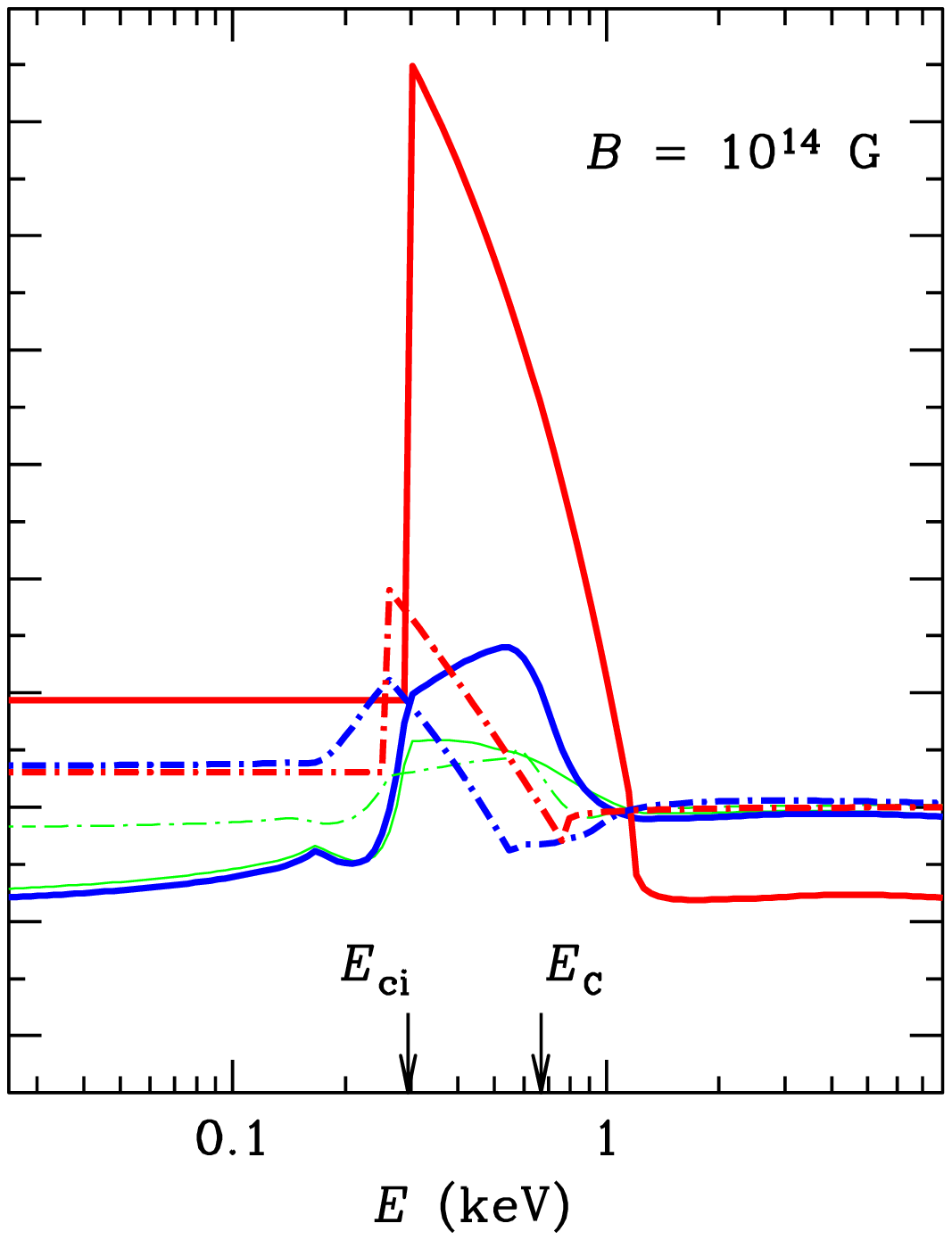}
\caption{Average linear polarization (normalized Stokes parameter
$Q/I$ in the reference frame where the $x$-axis is coplanar with the
magnetic axis) is shown as a function of photon energy for the hot spot
(red curves) or dipole field (blue curves), calculated in the model of
free ions for a condensed magnetized  iron surface, for two angles
between the line of sight and magnetic axis
$\theta_\mathrm{m}=90^\circ$ or $45^\circ$ (solid and dot-dashed
lines, respectively) and the magnetic field at the pole (or in the hot
spot) $B=10^{13}$~G (\emph{left panel}) or $10^{14}$~G (\emph{right
panel}). In the case of the dipole model, the results obtained without
allowance for the magnetospheric QED birefringence are shown by thin
green lines for comparison. Energies $E_\mathrm{ci}$, $E_\mathrm{C}$,
and $E_\mathrm{pe}$, marked at the horizontal axis, are the same as in
Fig.~\protect\ref{fig:rb}.
\label{fig:pol}
}
\end{center}
\end{figure}

After a photon has left the surface of a neutron star with a strong
magnetic field, it travels through the magnetosphere and experiences
the influence of vacuum polarization (see Sect.~\ref{sec:RT}), which
induces a change in the wave electric field as photon propagates. If
the magnetic field is sufficiently strong, then in the vicinity of the star
a typical length scale of this change is much shorter
than the scale of the magnetic field; in this case the photon propagates
adiabatically, so that its polarization instantaneously adapts to the
variation of the magnetic field direction
\cite{HeylShaviv00,HeylShaviv02}. Farther from the star the field
decreases, and eventually photons
leave the adiabatic region and maintain their polarization. The rays
that leave the adiabatic region pass through only a small solid
angle; consequently, polarizations of the rays
originating in different regions will tend to align together.
This effect partly lifts the strong suppression of the net
polarization caused by the averaging over the field and temperature
distributions at the star surface and can generate large net
polarizations \cite{HeilSL03}.

A comparative analysis of the polarization of radiation from
magnetized neutron stars with gaseous atmospheres and condensed
surfaces was recently performed in \cite{GonzalezCaniulef_ea16}. The authors
proposed a simplified treatment of photon propagation through the
adiabatic region. Instead of integrating the equations that govern the
photon polarization from the neutron star surface to the infinity, the
propagation is assumed perfectly adiabatic (i.e., each
photon preserves its polarization with respect to the local magnetic
field direction) up to some radius, where the typical scales of
variations of the photon electric field and external magnetic field
become comparable, and the change of polarization at larger
distances is neglected. This \emph{adiabatic radius} is estimated as
$r_\mathrm{ad}=4.8(B/10^{11}~\mbox{G})^{2/5}(E/\mbox{1 keV})^{1/5}R$.

Figure~\ref{fig:pol} shows the linear polarization degree in the model
of free ions for the two cases: first, where the radiation spectrum is
formed at a hot spot with $\bm{B}$ normal to the surface, and second,
where the field is distributed according to the relativistic dipole
model. In the latter case, an isotropic internal (core) temperature
of $10^8$~K is assumed and a surface temperature distribution is calculated
according to Ref.~\cite{PPP15} (this calculation gives
$\Teff\approx1.0$~MK for $B_\mathrm{p}=10^{13}$~G and
$\Teff\approx1.3$~MK for $B_\mathrm{p}=10^{14}$~G). The figure
illustrates the polarization smearing away by the averaging over the
stellar surface and the opposite effect due to the adiabatic
propagation of radiation in the near zone of the star.

\section{Theory versus observations}
\label{sec:obs}

As argued above,  models of strongly magnetized ($B\gg10^9$~G)
neutron-star atmospheres must take the bound species and their
radiative transitions into account. Let us consider a few examples
where models of magnetized, partially ionized atmospheres have been
used to study their thermal radiation.

\subsection{RX~J1856.5--3754}
\label{sec:1856}

RX~J1856.5$-$3754 is the closest and brightest of the class of X-ray
(dim) INSs (XINSs or XDINS, also known as the Magnificent Seven), whose X-ray
spectra are apparently of purely thermal nature (see \cite{Turolla09}
for a review). The spectra of the other XINSs show broad absorption
features at several hundred eV, and at least four of them also exhibit
narrow absorption features \cite{Hohle_ea12}. RX~J1856.5$-$3754 is the
only one of the Seven that has a blackbody-like X-ray spectrum without
absorption features. Similar to the other XINSs, its  X-ray and optical
spectra correspond to substantially different effective temperatures
if fitted separately with blackbodies. However, the measured spectrum
was fitted in the entire range from X-rays to optical within
observational errorbars with the use of the model of a thin magnetized
hydrogen atmosphere on top of a condensed iron surface \cite{Ho_ea07}.
The best agreement between the theoretical and observed spectra has
been achieved for the atmosphere column density $\ycol=1.2$ g
cm$^{-2}$, gravitational redshift $z_g\approx 0.25\pm0.05$, and apparent
radius $R_\infty=17.2^{+0.6}_{-0.1}\,D_{140}$  km, where
$D_{140}$ is distance in units of 140 pc. The authors used a
piecewise approximation to dipole-like distributions of the magnetic
field and temperature with the polar values
$B_\mathrm{p}\sim6\times10^{12}$~G, $\Ts\sim 0.7$~MK, whereas the
overall effective temperature is $\Teff^\infty=0.434\pm0.003$~MK
($1\sigma$ significance). Using an updated distance estimate,
$D=123^{+11}_{-15}$ pc \cite{Walter_ea10}, the above estimates of
$z_g$ and $R_\infty$ give $R=12.1^{+1.3}_{-1.6}$ km and
$M=1.5\pm0.3\,M_\odot$. Note that a fit of the observed X-ray spectrum
with the Planck function yields a 70\% higher temperature and a 3.5
times smaller radius of the emitting surface. Such huge difference
exposes the importance of a correct physical interpretation of an
observed spectrum for evaluation of neutron star parameters (as well
as in the non-magnetic thermally emitting INSs; cf.{}
Ref.~\cite{Heinke_ea14}).

The inferred magnetic field is significantly smaller than
$B_\mathrm{s}\approx1.5\times10^{13}$~G obtained from timing analysis
\cite{vKK08}. Here, $B_\mathrm{s}$ is so-called spindown or
characteristic magnetic field, that is the value of the equatorial
magnetic field in the ``canonical model'' of a rotating
non-relativistic magnetic dipole in vacuum \cite{Deutsch55}, with the
rotation axis orthogonal to the magnetic axis, with radius $R=10$ km,
and with moment of inertia $I=10^{45}$ g cm$^2$. Assuming different
$R$ and $I$, one would obtain another field estimate, proportional to
$\sqrt{I}/R^3$. For instance, modern BSk21 equation of state for the
neutron star matter (Ref.~\cite{Potekhin_ea13} and references therein)
yields $R=12.6$ km and $I=2.06\times10^{45}$ g cm$^2$ for
$M=1.5\,M_\odot$, which gives a factor 0.7 to the canonical
$B_\mathrm{s}$ estimate. Still more realistic estimate should take
into account that a real neutron star is surrounded by magnetosphere,
and the angle $\xi$ between its rotational and magnetic axes can
differ from $90^\circ$. An analytic fit to the results of numerical
simulations presented in Ref.~\cite{Spitkovsky06} gives a correction
factor $\approx0.8/(1+\sin^2\xi)$ to the canonical estimate of $B$.
Taken together, the two factors translate the result of
Ref.~\cite{vKK08} into $B\sim(4-8)\times10^{12}$~G. On the other
hand, an alternative interpretation of the spindown in terms of the
fallback disk model leads to still smaller estimate $B\sim10^{12}$~G
\cite{Ertan_ea14}.

Analysis \cite{Ho07} of the light curve of RX~J1856.5$-$3754 with the
aid of the same thin-atmosphere model yields constraints on the
geometry of rotational and magnetic axes. It turned out that the light
curve can be explained if one of these angles is small ($<6^\circ$),
while the other angle lies between  $20^\circ$ and $45^\circ$. In this
case, the radio emission around the magnetic poles does not cross the
line of sight. As noted in \cite{Ho07}, this may explain the
non-detection of this star as a radio pulsar \cite{Kondratiev_ea09}.

\subsection{RX J1308.6+2127}
\label{sect:RBS1223}

Hambaryan et al.~\cite{Hambaryan_ea11} described the
X-ray spectrum of RX J1308.6+2127 (RBS 1223) by a
wide absorption line centered around $\hbar\omega=0.3$~keV, superposed
on the Planck spectrum, with the line parameters depending on the
stellar rotation phase. This source shows the highest pulsed fraction
($13-42$\%, depending on energy band) of all the XINSs. A phase
resolved spectrum obtained from all 2003\,--\,2007 \textit{XMM-Newton}
observations of this star has been reproduced by the model a
magnetized iron surface covered by a partially ionized hydrogen
atmosphere with $\ycol\sim1$\,--\,10 g~cm$^{-2}$, with mutually
consistent asymmetric bipolar distributions of the magnetic field and
the temperature, with the polar values
$B_\mathrm{p1}=B_\mathrm{p2}=(0.86\pm0.02)\times10^{14}$~G,
$T_\mathrm{p1}=1.22^{+0.02}_{-0.05}$ MK, and
$T_\mathrm{p2}=1.15\pm0.04$ MK. The effective temperature is 
$\Teff\approx0.7$ MK. The gravitational redshift is estimated to be
$z_g=0.16^{+0.03}_{-0.01}$, which converts into
$(M/M_\odot)/R_6=0.87^{+0.13}_{-0.05}$ and suggests a stiff EOS of the
neutron star matter.

\subsection{1E~1207.4$-$5209}

The discovery of absorption lines
by \cite{Sanwal_ea02,Bignami_ea03} in the spectrum of CCO 1E
1207.4$-$5209 at energies $E\sim0.7\,N$~keV ($N=1,2,\ldots$)
immediately led to the belief that they are caused by
cyclotron harmonics. As we mentioned in
Sect.~\ref{sec:physinput}, such harmonics could be only electronic,
because the ion cyclotron harmonics are unobservable. Therefore, this
interpretation implies $B\approx7\times10^{10}$~G. Initially this interpretation
was criticized because the relative strengths of the spectral absorptions caused
by the cyclotron fundamental and
harmonics were thought to be sharply different, but this argument
was based on the considerations in the cold plasma approximation and
does not hold beyond it \cite{SuleimanovPW12} (as confirmed by
our Fig.~\ref{fig:spectra}).

There are doubts concerning the statistical significance of the third
and fourth observed lines \cite{MoriCH}, while the first and second
lines were tentatively interpreted as characteristic features in an
oxygen atmosphere with magnetic field $B\approx10^{12}$~G
\cite{MoriHailey06}. However, recent timing analysis
\cite{HalpernGotthelf15} gives the characteristic field
$B_\mathrm{s}=9.8\times10^{10}$~G and thus favors the
electron-cyclotron interpretation.

\subsection{2XMM J104608.7-594306}

2XMM J104608.7-594306 is a thermally emitting INS, but unlike the
Magnificent Seven it has a short spin period of only 18.6 ms. Its
\textit{XMM-Newton} spectrum has been analyzed in \cite{Pires_ea15},
using the blackbody model and hydrogen atmosphere models \textsc{nsa}
and \textsc{nsmaxg}. Statistically acceptable spectral fits and
meaningful physical parameters for the source are only obtained when
the purely thermal spectrum is modified by absorption lines at
$E\approx0.55$~keV and $1.3$~keV. The authors fixed $M$ to 1.4
$M_\odot$ and tried various values of magnetic field from 0 to
$3\times10^{13}$~G. In the case of the \textsc{nsmaxg} model, the best
statistical significance ($\chi^2/\mbox{d.o.f.}<1.1$) was provided by
the assumptions of $B=10^{10}$~G and $B=3\times10^{13}$~G. In the
former case, the inferred apparent radius $R_\infty=9$~km is
implausibly small, whereas in the latter case $R_\infty=12$~km does
not contradict to theoretical neutron star models. In the latter
case, the inferred effective temperature is
$\Teff=0.86^{+0.07}_{-0.04}$~MK (for comparison, the blackbody
model gives the redshifted temperature $\Teff^\infty\sim
(1.6-1.9)$~MK and $R_\infty\sim1.5-3$~km).

\subsection{1WGA J1952.2+2925}

1WGA J1952.2+2925 is the central X-ray source in the pulsar
wind nebula DA 495, but it shows no pulsar activity. An analysis
\cite{Karpova_ea15} of its archival \textit{Chandra} and
\textit{XMM-Newton} data shows that it has a pure thermal spectrum which is
equally well fitted either by the blackbody model with a temperature
of $\kB T \approx 215$ eV and an emitting area radius of $\approx 0.6$
km or by magnetized neutron star atmosphere model \textsc{nsmax} with
$\kB \Teff \sim 80-90$ eV. In the latter case the thermal emission can
come from the entire neutron star surface. The authors
\cite{Karpova_ea15} also placed an upper limit on the non-thermal flux
and an upper limit of 40\% on the pulsed fraction. The authors
suggested that the \textit{Fermi} source 3FGL J1951.6+2926 is the
likely $\gamma$-ray counterpart of 1WGA J1952.2+2925. 

\subsection{Rotation powered pulsars: PSR J1119$-$6127, B0943+10, 
J0357+3205, and J0633+0632}
\label{sect:pulsarfit}

The partially ionized, strongly magnetized hydrogen atmosphere model
\textsc{nsmax} (Sect.~\ref{sec:models}) was applied in \cite{Ng_ea12}
to interpret observations of pulsar J1119$-$6127, for which
the estimate based on spindown gives an atypically high characteristic
field $B_\mathrm{s}=4\times10^{13}$~G. In the X-ray range, it emits
pulsed radiation, with a significant thermal component. At fixed
$D=8.4$ kpc and $R=13$ km, the bolometric flux gives
$\Teff\approx1.1$~MK. It was difficult to explain, however, the large
pulsed fraction ($48\pm12$\%) by the thermal emission. Ng et
al.~\cite{Ng_ea12} managed to reproduce the X-ray light curve  of this
pulsar assuming that one of its magnetic poles is surrounded by an
area, which is covered by hydrogen and heated to $1.5$~MK, while the
temperature of the opposite polar cap is below $0.9$~MK.

A similar analysis was applied in \cite{Storch_ea14} to interpret
observations of pulsar B0943+10, which show correlated radio and
X-ray mode switches. The authors take $B=2\times10^{12}$~G
inferred from the pulsar spindown, assumed $M=1.2\,M_\odot$ and $R=12$
km, and modeled the emitting area as a hot spot covered by a partially
ionized hydrogen atmosphere. They find that an atmosphere with
$\Teff\approx(1.4-1.5)$~MK and emission radius
$R_\mathrm{em}\approx85$~m matches the radio-quiet X-ray spectrum,
whereas previous blackbody fits gave temperature $\approx3$~MK and
emission radius $\approx20-30$~m. The authors show that the large
X-ray pulse fraction observed during the radio quiet phase can be
explained by including the beaming effect of a magnetic atmosphere.

PSR J0357+3205 is an unusual thermally-emitting pulsar observed in
X-rays. Its spectrum was fitted in \cite{Kirichenko_ea14} with several
different multicomponent models. In the physically realistic case
where the incomplete ionization of the atmosphere was taken into
account, the authors used the \textsc{nsmax} model for the thermal
spectral component and a power-law model for the non-thermal one and
fixed $M=1.4\,M_\odot$ and $B=10^{12}$~G. They obtained an acceptable
fit ($\chi^2/\mbox{d.o.f}=1.05$) with a loose constraint on the
radius, $R_\infty=8^{+12}_{-5} (D/500\mbox{ pc})$~km.

Danilenko et al.~\cite{Danilenko_ea15} analyzed \textit{Chandra}
observations of the bright \textit{Fermi} pulsar J0633+0632 and confirmed early
findings that its X-ray spectrum contains non-thermal and thermal
components. The latter is equally well described by the blackbody
model and magnetized atmosphere model \textsc{nsmax} and can be
attributed to the emission from the bulk of the stellar surface. In
the latter case $\kB\Teff=53^{+12}_{-7}$~eV (at 90\% confidence),
which makes this pulsar  one of the coldest middle-aged INSs with
measured temperatures. The authors also reported evidence of an
absorption feature at $E=804^{+42}_{-26}$ eV with equivalent width of
$63^{+47}_{-36}$ eV.

\section{Conclusions}
\label{sec:concl}

We have considered the main features of neutron-star atmospheres and
radiating surfaces and outlined the current state of the theory of the
formation of their spectra
(see \cite{P14} for more details).
The interpretation of observations enters
a qualitatively new phase, free from the assumptions of a blackbody
spectrum or the ``canonical model'' of neutron stars. Spectral
features, compatible with absorption lines in some cases, have been
discovered in thermal spectra of strongly magnetized neutron stars. On
the agenda is their detailed theoretical description, which may
provide information on the surface composition, temperature and
magnetic field distributions. However, in order to disentangle these
parameters, a number of  problems related to the theory of magnetic
atmospheres and radiating surfaces  still have to be solved.

\emph{Acknowledgements}.
A.Y.P.{} is grateful to the organizers of the conference  ``The Modern
Physics of Compact Stars 2015'' for the invitation, hospitality, and
financial support of participation.  We thank the referee, Valery
Suleimanov for careful reading the manuscript and useful remarks. The
work of A.Y.P.{} has been partly  supported  by the RFBR  (grant
14-02-00868). W.C.G.H. appreciates use of computer facilities at
KIPAC.


\end{document}